\documentstyle[subeqn,graphicx,natbib,txfonts]{aa}
\bibpunct{(}{)}{;}{a}{}{,}

\input epsf
\newcommand{\beq}{\begin{equation}}
\newcommand{\eeq}{\end{equation}}
\newcommand{\ben}{\begin{enumerate}}
\newcommand{\een}{\end{enumerate}}
\newcommand{\bei}{\begin{itemize}}
\newcommand{\eei}{\end{itemize}}
\newcommand{\bfig}{\begin{figure}}
\newcommand{\efig}{\end{figure}}
\newcommand{\bea}{\begin{eqnarray}}
\newcommand{\eea}{\end{eqnarray}}
\newcommand{\al}{\alpha}
\newcommand{\bt}{\beta}
\newcommand{\gm}{\gamma}

\newcommand{\dl}{\delta}
\newcommand{\Dl}{\Delta}

\newcommand{\lm}{\lambda}

\newcommand{\om}{\omega}

\newcommand{\rarrow}{\rightarrow}
\newcommand{\nn}{\nonumber}

\newcommand{\sigmaT}{\sigma_{\rm T}}

\newcommand{\gmin}{\gamma_{\rm min}}
\newcommand{\gmax}{\gamma_{\rm max}}
\newcommand{\melec}{m_{\rm e}}
\newcommand{\diff}{{\rm d}}

\newcommand{\epso}{\epsilon_{\rm 0}}
\newcommand{\epsmin}{\epsilon_{\rm 0, min}}
\newcommand{\epsmax}{\epsilon_{\rm 0, max}}
\newcommand{\bis}{b_{\rm is}}
\newcommand{\bes}{b_{\rm es}}
\newcommand{\besc}{b_{\rm esc}}
\begin{document}
\title{Spectral evolution of non-thermal electron distributions
in intense radiation fields}
\author{ K. Manolakou$^1$, D. Horns$^1$ \and J.G.Kirk$^2$
\institute{ $^1$ Institute for Astronomy and Astrophysics T\"ubingen (IAAT)
Sand~1, D-72076 T\"ubingen, Germany\\
$^2$ Max-Planck-Institut f\"ur Kernphysik, Saupfercheckweg~1, D-69117 Heidelberg, Germany}
}
\offprints{horns@astro.uni-tuebingen.de}
\date{Received / Accepted}
\titlerunning{Non-thermal electrons in radiation fields}
\authorrunning{Manolakou, Horns \& Kirk}
\abstract  
{Models of many astrophysical gamma-ray sources assume they contain a
  homogeneous distribution of electrons that are injected as a
  power-law in energy and evolve by interacting with radiation fields,
  magnetic fields and particles in the source and by escaping. This
  problem is particularly complicated if the radiation fields have
  higher energy density than the magnetic field and are sufficiently
  energetic that inverse Compton scattering is not limited to the
  Thomson regime.}
{We present a simple, time-dependent, 
semi-analytical solution of the electron kinetic 
equation that treats both continuous and impulsive injection, 
cooling via synchrotron and
  inverse Compton radiation, (taking into account Klein-Nishina effects)
  and energy dependent particle escape. We use this solution to calculate 
  the temporal
  evolution of the multi-wavelength spectrum of systems where 
energetic electrons cool in intense photon fields.  }
{ The kinetic equation for an arbitrary, time-dependent source
  function is solved by the method of Laplace transformations. Using
  an approximate expression for the energy loss rate that takes into
  account synchrotron and inverse Compton losses including
  Klein-Nishina effects for scattering off an isotropic photon field
  with either a power-law or black-body distribution, we find
  explicit expressions for the cooling time and escape probability of
  individual electrons. This enables the full, time-dependent solution
  to be reduced to a single quadrature. From the electron
  distribution, we then construct the time-dependent, multi-wavelength
  emission spectrum. }
{ We compare our solutions with several limiting cases and discuss the
 general appearance and temporal behaviour of spectral features 
 (i.e., cooling breaks, bumps etc.). As a specific example, we model the 
  broad-band energy spectrum of the open stellar association Westerlund-2
  at different times of its evolution, and compare it with
  observations.} 
{The technique we present enables simple, computationally efficient,
  time-dependent models of homogeneous sources to be constructed and compared
  with multi-wavelength observations.}

\keywords{Gamma rays: theory -- Radiation mechanisms: non-thermal -- Stars: Wolf-Rayet -- (ISM) cosmic rays -- open clusters and associations: individual: Westerlund 2 }
\maketitle
\section{Introduction}
The discovery of very high energy (VHE, $E_\gamma>100$~GeV)
gamma-rays from many different astrophysical sources 
\citep[for a recent review see][]{2006IAUS..230...95V}
has motivated the construction of simple, physical 
models of the emission regions. 
An important ingredient in such models is the interaction of 
accelerated leptons with an intense ambient radiation field, 
especially by inverse Compton scattering. For example,
the recent discovery of VHE emission from 
sources possibly associated with stellar clusters
including Westerlund-2 (Wd-2) \citep{2007A&A...467.1075A}, Berk-87
\citep{2007ApJ...658L..33A}, and Cyg~OB2
\citep{2002A&A...393L..37A,2005A&A...431..197A} provides evidence for
particle acceleration in an environment dominated by hot, massive
stars driving fast winds, as well as supernova remnants evolving in a
dilute, hot medium \citep{2005ApJ...628..205T}.  
The detection of
VHE gamma-rays from high mass X-ray binary systems like LS I+61\,303
\citep{2006Sci...312.1771A}, LS~5039 \citep{2005Sci...309..746A}, and
PSR~B1259-63 \citep{2005A&A...442....1A}
adds another class of galactic sources where  
inverse Compton scattering is expected to be an important energy loss 
mechanism.

In addition to the
VHE emission, these sources show evidence for non-thermal X-ray emission, 
which, in the case of Wd-2 and Cyg~OB2, is also spatially extended
\citep{2005xrrc.procE3.04T,2007arXiv0705.0009H}. 
Leptonic scenarios have been used both to predict and to model the 
multi-wavelength spectra 
\citep{1999APh....10...31K,2003PASJ...55..473M,2007Ap&SS.tmp..212K,2007ApJ...657..302H}, although hadronic models in which the gamma-rays are produced in 
nucleonic collisions also provide a natural explanation
\citep{2006astro.ph.10139N,2007arXiv0705.0009H,2007arXiv0704.3517B}.


Improved modelling of these sources is clearly necessary in order to 
reach a better understanding of the mechanisms at work.
However, even with drastic 
simplifications, such as the assumption of spatial homogeneity, 
and the {\em ad hoc} prescription of a distribution function of injected particles,
such models can be quite complicated to construct. Furthermore, 
the need to investigate a large volume of parameter space makes 
it important to find computationally efficient algorithms.  

The temporal evolution of a non-thermal electron distribution
interacting with matter and photon fields has been studied in great
detail in the past (see e.g.
\citet{1964ocr..book.....G,1966ApJ...146..686F,1970RvMP...42..237B,1997A&A...320...19M,1997ApJ...490..619S}).
The problem can be described by an integro-differential equation
including all relevant energy loss mechanisms for electrons (e.g.
\citet{1970RvMP...42..237B,1989ApJ...342.1108Z,1992MNRAS.258..657C,2005ApJ...621..285K})
which is however numerically quite difficult to treat. 
The problem
can be simplified considerably by approximating the 
discrete jumps in energy suffered by an electron undergoing
Compton 
scatterings as a continuous energy
loss process. In this case, the kinetic equation reduces to a partial 
differential equation. This approximation, which is accurate in the 
Thomson limit,  appears to be reasonably good 
also in the Klein-Nishina regime for a wide range of target photon
distributions \citep[see, e.g.,][]{1989ApJ...342.1108Z}.

Analytical solutions to the electron kinetic equation 
have been found for special cases
e.g., when continuous or impulsive injection is considered, with the
electrons cooling radiatively through synchrotron and inverse Compton
scattering
in the Thomson limit as well as escaping from the system \citep{1962AZh....39..393K}.
Although Klein-Nishina (KN) effects are well-known to have 
an interesting impact on the steady state spectra
\citep[e.g.,][]{1971PhRvD...3.2308B,1999APh....10...31K,2002ApJ...568L..81D,2005MNRAS.363..954M}
their inclusion in time-dependent models has so far necessitated 
a rather elaborate numerical treatment 
\citep{1997A&A...320...19M,2002MNRAS.336..721K}.
However, recently, a
useful approximation for the treatment of the inverse Compton energy
losses in an isotropic photon field, 
including the transition from Thomson to the KN regime has been
presented \citep{2005MNRAS.364.1488M}.
In this paper, we use this approximation to develop a
semi-analytical solution for the electron kinetic equation.  The
numerical treatment of the time-dependent problem is thus reduced to a
single quadrature.  We use this method to discuss the temporal
evolution of characteristic features in the electron distribution as a
consequence of energy losses in the KN regime.  These effects are of
interest in the sources mentioned above, as well as in active galactic
nuclei.  The paper is structured in the following way: in
\S\ref{sec:elec}, we derive our semi-analytical solution to the continuity
equation describing the temporal evolution of an electron distribution
suffering synchrotron, inverse Compton (including KN effects), and
escape losses for an arbitrary, time-dependent injection
spectrum.  The necessary analytical calculations are summarized in 
Appendices~A and B.  Numerical results found using this method
are presented in \S\ref{sec:numerical} and a specific
application to the case of the stellar
association Wd-2 is given in \S\ref{sec:applic}. We close the paper in
\S\ref{sec:conclusions} with a discussion of the results and comment on
further refinements of the calculations.
\section{Electron energy distribution}
\label{sec:elec}
\subsection{Formal solution of the kinetic equation}
Assume a homogeneous source into which an unspecified acceleration
process injects electrons at the rate $Q(\gm,t)$, which subsequently
cool and escape.  The kinetic equation governing the evolution of 
the number $N(\gm,t)$ of electrons in the source with
Lorentz factors $\gm$ in the interval $\diff \gm$ at time $t$ is: 
\bea 
{\partial N(\gm,t) \over \partial t} &=&
{\partial \over \partial \gm} \{\dot{\gm}(\gm) N(\gm,t)\} - \nu_{\rm esc}(\gm)
N(\gm,t)+Q(\gm,t) \enspace.
\label{parN}
\eea 
In writing Eq.~(\ref{parN}) we have assumed that each individual
electron in the source loses energy continuously, according to $\diff
\gm / \diff t =-\dot{\gm}$ (note that $\dot{\gm}$ is defined to be positive) 
and has an energy dependent probability
$\nu_{\rm esc}(\gm)\diff t$ of escaping the source in a time interval
$\diff t$.  The continuous-loss approximation is appropriate for
synchrotron radiation and for Compton scattering in the Thomson limit,
as well as for collisional losses (\lq\lq ionisation losses\rq\rq) in
a fully ionized plasma. In the case of Compton scattering in the
Klein-Nishina regime, this approximation has been examined by several
authors \citep[e.g.,][]{1971PhRvD...3.2308B,1989ApJ...342.1108Z}. Provided one
is interested only in spectral features that appear in the transition
region between the Thomson limit and the extreme Klein-Nishina limit,
it appears to be justified. 
In addition, we assume that Eq.~(\ref{parN}) is linear in $N(\gm,t)$,
i.e., that the the emitted radiation has no further effect on the
electron distribution, or on the electron injection function. This
excludes situations encountered in compact sources in which feedback
effects can be important
\citep[e.g.,][]{2006A&A...451..739K,1987ApJ...319..643L,2007ApJ...661L..17S},
but is not a significant restriction for objects such as stellar clusters.

In the case discussed here, where the total energy-loss rate,
$\dot{\gm}(\gm)$, and the escape rate, $\nu_{\rm esc}(\gm)$, are
independent of time, Eq.~(\ref{parN}) can be solved by several
standard methods, e.g., by
means of the Green's function for the homogeneous equation 
\citep{1964ocr..book.....G}, or using Laplace transforms 
\citep{1980gbs..bookR....M}.
Following \citet{1980gbs..bookR....M}, we define the 
quantity $\tau(\gm',\gm)$ that is the time required for an electron 
to cool from a Lorentz factor of $\gm'$ to one of $\gm$ ($\le\gm'$):
\beq
\tau(\gm',\gm):= \int_\gm^{\gm'}\!\!\frac {\diff \gm''}{\dot{\gm}(\gm'')}
\label{taudefinition}
\eeq
and the quantity $\lm(\gm',\gm)$ such that 
$1-\exp\left[-\lm\left(\gm',\gm\right)\right]$ 
is the probability that an electron escapes whilst cooling from $\gm'$ to $\gm$:
\beq
\lm(\gm',\gm):= \int_\gm^{\gm'}\!\!
\diff {\gm''}\frac {\nu_{\rm esc}(\gm'')}{\dot{\gm}(\gm'')}
\enspace.
\label{f1f2}
\eeq
This enables the solution to be written:
\bea
\lefteqn{N(\gm, t)\,=\,\frac{1}{\dot{\gm}(\gm)}\;\;\int_\gm^\infty
\!\!\diff \gm'\textrm{e}^{-\lm(\gm',\gm)}\,\times\, }&&
\label{solN}
\\ 
&&\!\!\!
\left\lbrace Q\left[\gm',t-\tau(\gm',\gm)\right]
H \left[t-\tau(\gm',\gm)\right]
+ N(\gm',0)\dl\left(t-\tau(\gm',\gm)\right)\right\rbrace,
\nonumber
\eea
where $H$ and $\dl$ stand for the Heaviside and delta functions, respectively.
On multiplying Eq.~(\ref{solN}) by $\dot{\gm}(\gm)$, it is
straightforward to assign a physical interpretation to each
term in this solution. Thus, 
\begin{description}
\item
$\dot{\gm}(\gm)N(\gm,t)$ 
is the number of particles per
unit time that cool across the Lorentz factor $\gm$,
\item
$\textrm{e}^{-\lm(\gm',\gm)} Q\left[\gm',t-\tau(\gm',\gm)\right]H
\left[t-\tau(\gm',\gm)\right]\diff\gm'$ is the rate at which particles injected
with Lorentz factor between $\gm'$ and $\gm'+\diff\gm'$
arrive at time $t$ at a Lorentz factor $\gm$, and
\item
$\textrm{e}^{-\lm(\gm',\gm)}
N(\gm',0)\dl\left(t-\tau(\gm',\gm)\right)\diff\gm'$ is the 
rate at which particles whose initial Lorentz factor lies
between $\gm'$ and $\gm'+\diff\gm'$ arrive at Lorentz factor $\gm$
at time $t$.
\end{description}

As we shall see below, the cooling rate is such that in the limit of
large Lorentz factor, synchrotron losses always dominate:
$\dot{\gm}\sim\gm^2$ as $\gm\rightarrow\infty$. In this case, a finite
time suffices for an electron 
to cool down to $\gm$ from an initially arbitrarily large Lorentz
factor. Defining the function $\gm_1(t)$ as
the Lorentz factor at time $t$ of a particle
whose initial energy was infinite, i.e.,
\bea
\tau[\infty,\gm_1(t)]&=&t
\label{gamma01b}
\eea
one can rewrite Eq~(\ref{solN}) as
\bea
N(\gm, t) &=& 
\frac{1}{\dot{\gm}(\gm)} 
\Bigl\{\int\limits_\gm^{\gm_0}\!\!\diff \gm'\textrm{e}^{-\lm(\gm',\gm)}
\nn \\
\lefteqn{Q \bigl[\gm',t-\tau(\gm',\gm)\bigr]H\bigl[t-\tau(\gm',\gm)\bigr]
+ N(\gm_0,0)\textrm{e}^{-\lm(\gm_0,\gm)} \Bigr\} 
,}&&
\nn \\
&&\textrm{\ for\ }\gm\le\gm_1(t)
\label{sol2N}
\eea 
and 
\bea
N(\gm, t) &=& 
\frac{1}{\dot{\gm}(\gm)} 
\int\limits_\gm^{\infty}\!\!\diff \gm'\textrm{e}^{-\lm(\gm',\gm)}
Q \bigl(\gm',t-\tau(\gm',\gm)\bigr)  
,
\nonumber\\
&&\textrm{\ for\ }\gm>\gm_1(t)
\label{sol2N2}
\eea 
where $\gm_0(\gm,t)$ is defined (for $\gm\le\gm_1(t)$) by the relation
\bea
\tau(\gm_0,\gm)&=&t
\label{gamma01a}
\eea 

The exact functional form of Eqs.~(\ref{sol2N}) and (\ref{sol2N2}) 
is determined by the energy dependence of the 
radiation processes involved and the escape
rate, as well as the injection function specified.
As an illustration, we present the
well-known case of constant injection of a power-law spectrum: $Q(\gm,
t)= Q_0 \gm^{-p}$, starting at $t=0$ with $N(\gm,0)=0$, where
radiative cooling is due only to synchrotron emission:
$\dot{\gm}=b_s \gm^2$ and no escape term is present ($\nu_{\rm
  esc}=0$). In this case $\tau(\gm',\gm) =\left(b_s\gm'\right)^{-1}
-\left(b_s \gm\right)^{-1}$ and
$\lm(\gm',\gm)=0$. Eq.~(\ref{sol2N}) then gives (see also
\citet{1980gbs..bookR....M}): \beq N(\gm,t)=\frac{Q_0\gm^{-p}}{(p-1)
  b_s \gm} \left\lbrace \begin{array}{ll}
    [1-(1-b_s\gm t)^{p-1}]:& 0\le b_s\gm t \le 1  \\
    &       \\
    1: & b_s\gm t > 1
\end{array}
\right.
\label{egsol}
\eeq
As implied by Eq.~(\ref{egsol}) the electron spectrum steepens from 
$\gm^{-p}$ for $b_s\gm t \ll 1$ to
$\gm^{-(p+1)}$ for $b_s\gm t \approx 1$ with a break at $\gm \approx 1/b_s t
=\gm_1(t)$ that moves to lower energies
as time progresses: the well-known \lq\lq spectral ageing\rq\rq\ effect.

For an injection function  which is non-zero only within a
range of the Lorentz factor $\gm \in [\gmin,\gmax]$ 
\beq
Q=
\left\lbrace \begin{array}{ll}
Q(\gm,t)& \textrm{for $\gmin \le \gamma \le \gmax$ } \\
        &       \\
0 & \textrm{otherwise}
\end{array}
\right.
\label{inj1}
\eeq
it is 
computationally  more efficient to take account of this explicitly
by modifying the integration limits in Eqs.~(\ref{sol2N}) and
(\ref{sol2N2}). Specialising to 
$N(\gm,0)=0$, one finds:
\bea
N(\gm, t)&=& 
\frac{1}{\dot{\gm}(\gm)}\;\;\int\limits_{\gm_l}^{\gm_u}\!\!\diff \gm'
\textrm{e}^{-\lm(\gm',\gm)}
Q \bigl[\gm',t-\tau(\gm',\gm)\bigr]H \bigl[t-\tau(\gm',\gm)\bigr],
\nn\\
&&
\textrm{for $\gm_2 \le \gm \le \gmax$ } 
\label{sol3N}
\eea
\noindent
with $N(\gm, t)= 0$ for $\gm <\gm_2$ and $\gm>\gmax$.
Here, the lower and upper integration limits are:
\beq
(\gm_l^{\gmin<\gm_1}, \gm_u^{\gmin<\gm_1})=
\left\lbrace \begin{array}{ll}
(\gmin,\gm_0): & \gm_2 < \gm \le \gmin \\
(\gm,\gm_0): & \gmin < \gm \le \gm_1 \\
(\gm,\gmax): & \gm_1 < \gm \le \gmax, 
\end{array}
\right.     
\label{limits}
\eeq
\noindent 
when $\gmin<\gm_1$, while
\beq
(\gm_l^{\gm_1<\gmin},\gm_u^{\gm_1<\gmin})=
\left\lbrace \begin{array}{ll}
(\gmin,\gm_0): & \gm_2 < \gm \le \gm_1 \\
(\gmin,\gmax): & \gm_1 < \gm \le \gmin \\
(\gm,\gmax): & \gmin < \gm \le \gmax, 
\end{array}
\right.     
\eeq
\noindent 
for $\gm_1<\gmin$.\\
The value of $\gm_0$ fulfill still Eq.~(\ref{gamma01a}), but $\gm_1$
and $\gm_2$ are now defined by: 
\beq
\tau(\gmax,\gm_1)- t = 0,
\label{gamma1}
\eeq
\beq
\tau(\gmin,\gm_2)- t = 0.
\label{gamma2}
\eeq Eqs~(\ref{gamma1}) and (\ref{gamma2}) describe the two breaks,
$\gm_1$ and $\gm_2$ introduced in the
evolved spectrum due to the existence in the injection spectrum of a 
minimum and a maximum Lorentz factor, $\gmin$ and $\gmax$, 
respectively. $\gm_2 < \gm_1$ with both of them moving to lower 
energies as the time evolves. 

As a further illustration, Appendix~\ref{appendixb} presents the solution of 
Eq.~(\ref{parN}) for a power-law
type injection function with finite energy spectrum, where synchrotron
and inverse Compton losses in the Thomson regime are present and
electrons escape at a rate $\nu_{\rm esc} \propto \gm$ (see
\S~\ref{subrates}).

\subsection{Energy loss and escape rates}
\label{subrates}
In the continuous energy-loss approximation, 
the total energy loss rate of a relativistic
electron ($\gm \gg 1$), averaged over an isotropic distribution 
and taking into account synchrotron cooling,
inverse Compton scattering of the electrons on ambient photons,
Coulomb losses and Bremsstrahlung emission (in a fully ionised
hydrogen gas), is given by the formula: 
\bea
\dot{\gm}_{\rm tot} &=& b_s\gm^2+b_{\rm iC}\gm^2 F_{\rm KN}(\gm) 
\nn \\
&&+ b_C(\ln \gm+b_C^0)+b_B\gm(\ln \gm +b_B^0).
\label{gammadottotal}
\eea
\noindent
The coefficients, $b_s,$ $b_{\rm iC},$ $b_C,$ and $b_B$, and the 
constants $b_C^0$ and $b_B^0$ are given by the following relations 
\citep{1964ocr..book.....G}:
\begin{subequations}
\beq
b_s=\frac{4\sigmaT}{3 \melec c}u_{\rm B}=1.292\times 10^{-15} 
(B/{\rm mG})^2 \;\; \textrm{sec$^{-1}$}, 
\label{bs}
\eeq
\beq
b_{\rm iC}=b_s \frac{u_0}{u_{\rm B}}=5.204\times 10^{-20} (u_0/
{\rm eV~cm}^{-3})\;\; \textrm{sec$^{-1}$},
\label{bic}
\eeq
\beq
b_C=\frac{2 \pi e^4 n_e}{\melec^2 c^3}=1.491 \times 10^{-14} n_e 
\;\; \textrm{sec$^{-1}$}, 
\label{bc}
\eeq
\beq
b_B=\frac{4 e^6 n_e}{\melec^2 c^4 \hbar}=1.37 \times 10^{-16}n_e  \;\; 
\textrm{sec$^{-1}$},
\label{bb}
\eeq
\end{subequations}
\noindent
and 
\begin{subequations}
\beq
b_C^0=\ln\left(\frac{\melec^3 c^4}{4e^2 n_e \hbar^2}\right)+\frac{3}{4}=-\ln n_e +73.4, 
\label{bC0}
\eeq
\beq
 b_B^0=\ln2-\frac{1}{3}=0.36
\label{bb0}
\eeq
\end{subequations}
\noindent
Here, $\sigmaT$ is the Thomson cross-section, $u_0$, $u_{\rm B}$ are
the total energy densities of the photon and magnetic fields,
respectively, while $c$, $\melec$, and $e$ are the velocity of light
in vacuum, the electron mass and charge, respectively. Finally, $n_{\rm e}$
is the electron density in units of $\textrm{cm}^{-3}$.
In the case of ionization or Bremsstrahlung losses (in neutral 
hydrogen), the corresponding terms in Eq.~(\ref{gammadottotal}) 
should be replaced by $b_C(3 \ln \gm +18.8)$ and 
$b_B\left[\ln(191)+1/18\right]\simeq 5.3~b_B$.

The escape of particles from the source may be modeled in
different ways. The simplest method is to assume that the 
\lq\lq source\rq\rq\ is a 
region of relatively strong magnetic field and high target 
photon density. When particles
are carried out of this region by the motion
of the background plasma they may not only leave the target fields behind, 
but may also undergo a sudden drop in energy due to adiabatic expansion. 
This scenario implies that the escape rate
is independent of energy, being the reciprocal of the 
average time taken by a fluid element to
cross the region at a given speed. 

An alternative method, (which, however, does not take account of a possible 
energy loss) is to consider the particle transport as a diffusive
process with the spatial diffusion coefficient $D\equiv \langle \Dl
r^2\rangle /2 t$, being either constant or
a function of the energy of the particle or/and time, $D(\gm,t)$.
A common practice is to calculate the spatial diffusion coefficient in
the Bohm diffusion limit, which is defined by setting the step size of
the random walk of a particle to be equal to the Larmor radius,
$r_L=\gm \melec c/eB$ (for relativistic electrons). In this case,
$D(\gm) =r_L c /3$ and the escape rate {from
a spherical region with radial extension $R$}, $\nu_{\rm esc}(\gm)= 2
D(\gm)/R^2$ is, therefore, directly proportional to the particle's
energy: 
\beq \nu_{\rm esc}(\gm) = b_{\rm esc} \gm,
\label{escrate}
\eeq
\noindent
where 
$b_{\rm esc}=2 \melec c^2 /3 eB R^2 = 
3.6\times 10^{-21}~{\rm s}^{-1}~(B/{\rm mG})^{-1}\cdot (R/{\rm pc})^{-2}$.

\subsection{Inverse Compton scattering}

The function $F_{\rm KN}(\gm)$ in Eq.~(\ref{gammadottotal}) takes into
account the full Klein-Nishina cross section for Compton
scattering and is given by the relation: 
\beq 
F_{\rm  KN}(\gm)=
\frac{1}{u_0} \int\limits_0^{\infty} f_{\rm
  KN}(\gm,\epso)u_{\epso} \diff \epso,
\label{FKN}
\eeq
where $u_{\epso}$ is the differential photon energy density:
$u_0=\int u_{\epso}\diff\epso$ and $\epso=h\nu/mc^2$ is the dimensionless 
energy of a target photon of frequency $\nu$. The 
kernel, $f_{\rm KN}(\gm,\epso)$,
was first given by \citet{1965PhRv..137.1306J}.
It is important to note that this kernel, although originally derived for
an isotropic field of target photons, is also valid for an anisotropic 
photon field, provided the average over an isotropic {\em electron} 
distribution is taken \citep[see][Appendix~A]{1999APh....10...31K}.
A useful approximate expression in the limit $\gm\gg\epso$, 
has recently been given by \citet{2005MNRAS.364.1488M}:
\bea
f_{\rm KN}(\gm,\epso)&\approx&\left(1+4\gm\epso\right)^{-3/2},
\label{fKNapprox}
\\
\noalign{\hbox{for}}
\gm\epso\la 10^4
\nonumber
\eea
and these authors also estimate the implications of Eq.~(\ref{fKNapprox}) 
for the functional form of
$F_{\rm KN}(\gm)$, for different photon energy distributions.

For a Planckian distribution of photon energies, the Compton energy
losses for electrons are dominated by scatterings on photons with
dimensionless energy 
$\epsilon_{\rm eff}=2.8kT/\melec c^2$. For $kT\ll\gm\melec c^2$, 
the function $F_{\rm KN}(\gm)$ can
therefore be approximated by treating the Planckian as a
mono-energetic photon distribution (i.e., $u_{\epso}=a_0
\dl(\epso-\epsilon_{\rm eff})$).
Using this approximation and Eq.~(\ref{fKNapprox}) 
one finds:
\bea
F_{\rm KN}(\gm)&\approx& (1+4\gm\epsilon_{\rm eff})^\bt
\label{FKNapprox}
\\
\noalign{\hbox{with}}
\epsilon_{\rm eff}&=&2.8kT/\melec c^2 
\nonumber\\
\noalign{\hbox{and}}
\bt&=&-3/2.
\nonumber
\eea

For a power-law radiation field ($u_{\epso}=a_0 \epso^{-\al_0}$), 
$F_{\rm KN}(\gm)$ can be written in the same general form
as in Eq.~(\ref{FKNapprox}), provided 
the values of the parameters $\epsilon_{\rm eff}$ and $\bt$
are suitably chosen. 
For $\al_0<-0.5$ and
$\al_0>1$, inverse Compton losses in the KN limit are dominated by
scatterings on photons with the highest ($\approx \epsmax$) and the
lowest ($\approx \epsmin$) photon energies available, respectively.
One may therefore write, $\epsilon_{\rm eff}=\epsmax$ for 
$\al_0<-0.5$ and $\epsilon_{\rm eff}=\epsmin$
for $\al_0>1$, respectively, with $\bt=-3/2$ in each case.

In the intermediate case of $-0.5 < \al_0 < 1$ the relevant photon
energy range for scattering is broader ($\approx 1/4 \gm $) and one
may use instead the so-called \lq\lq Thomson edge\rq\rq\ or \lq\lq KN
cut-off\rq\rq\ approximation. In this range
$F_{\rm KN}(\gm)$ is
still given by Eq.~(\ref{FKNapprox}) with $\epsilon_{\rm eff} =\epsmax$
and $\bt =\al_0-1$, provided $\epsmin/\epsmax \ll
1$. Table~\ref{table1} summarizes the values of the parameters 
$\epsilon_{\rm eff}$ and $\bt$ for the case of a black-body photon targets, 
as well as for power-law target photon fields.

\bfig
\includegraphics[width=\linewidth]{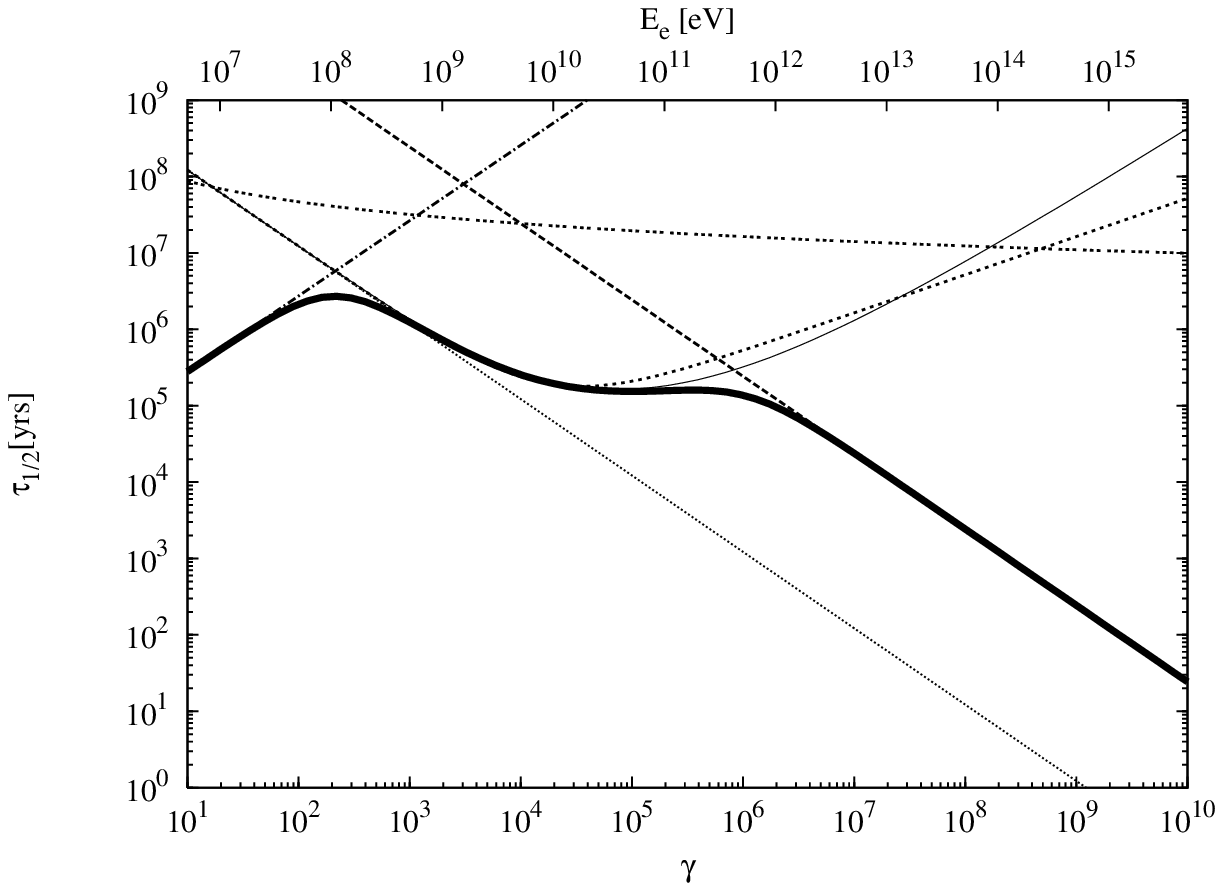}
\caption{\label{fig1} 
The cooling time $\tau_{\rm cool}=\gamma/\dot \gamma$
as a function of Lorentz factor  
for the various energy loss mechanisms for electrons
including Coulomb scattering (dash-dotted), Bremsstrahlung 
(triple dotted), synchrotron (long dashed), inverse Compton 
(Thomson limit: dotted,
approximation (see Eqn.\ref{FKNapprox}): short dashed, 
no approximation: thin line), and total (heavy line). The parameters
are $n_{\rm e}=1$~cm$^{-3}$, $u_0=500$~eV~cm$^{-3}$, 
$T=30\,000$~K, and $B=10~\mu$G, as expected 
for a stellar environment with hot young stars.}
\efig
\bfig
\includegraphics[width=\linewidth]{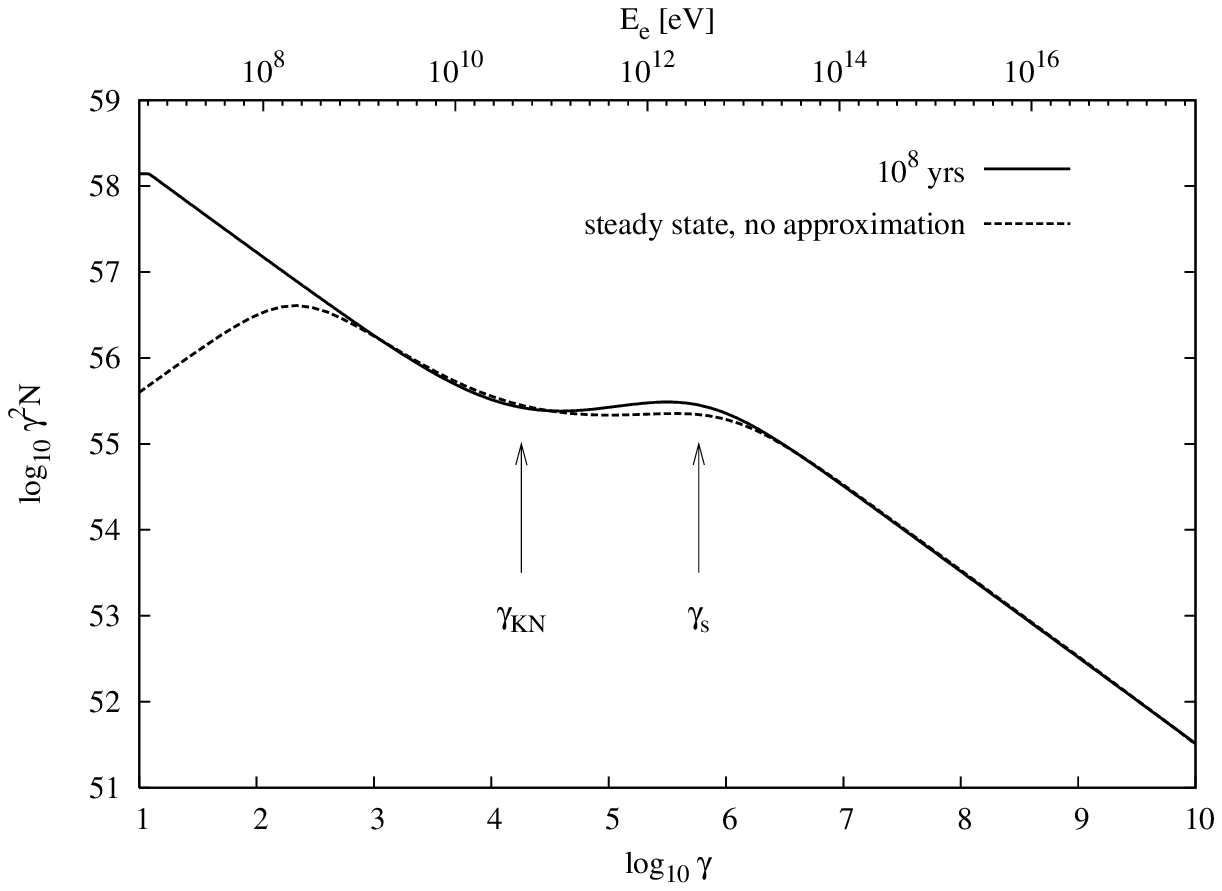}
\caption{\label{fig2}
 The solution $N(\gamma,t=10^{8}~{\rm yrs})$ using
the approximate cooling term of Eq.~(\ref{gammadotgen}) 
in an environment identical to the one used in Fig.~\ref{fig1} 
(solid line) compared with 
 the steady state solution using the full expression,
for $F_{\rm KN}$ and including Coulomb losses (dashed line).}
\efig
\begin{figure*}
\includegraphics[width=0.49\linewidth]{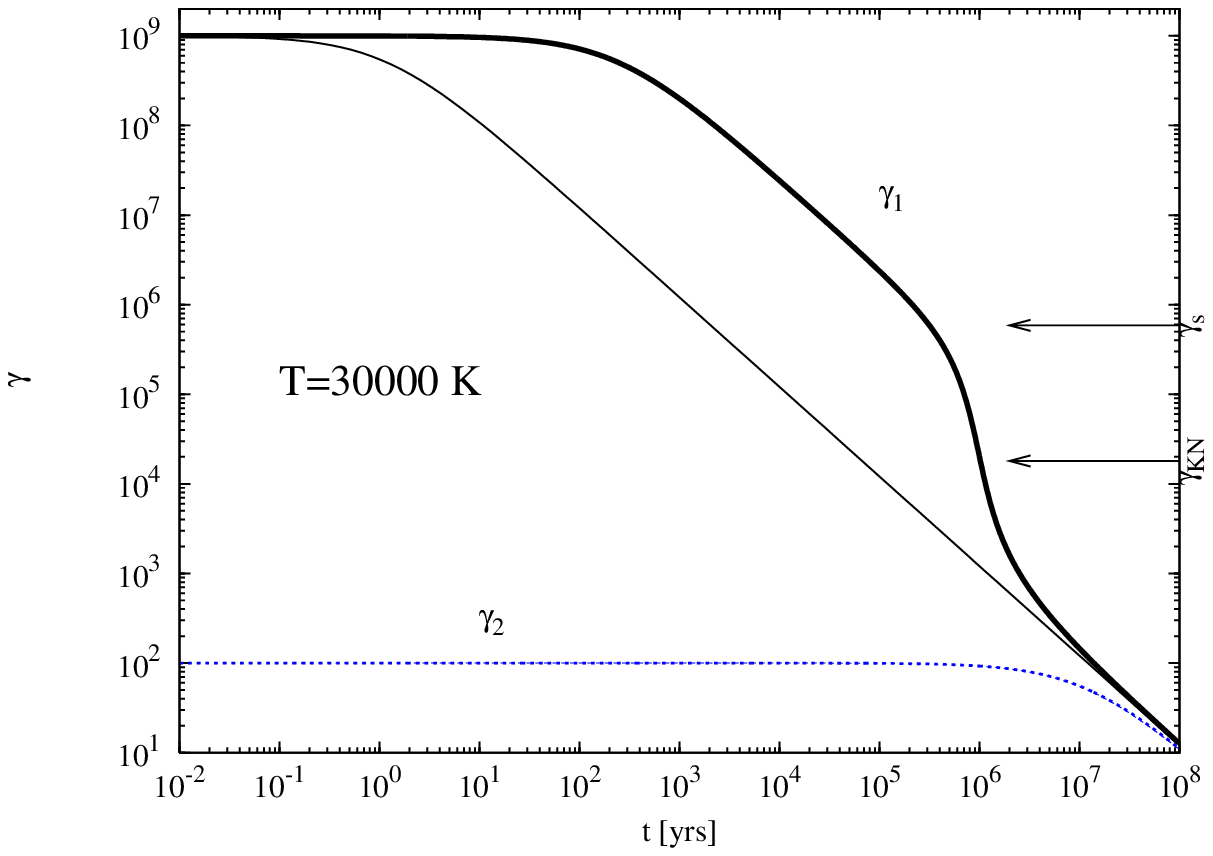}
\includegraphics[width=0.49\linewidth]{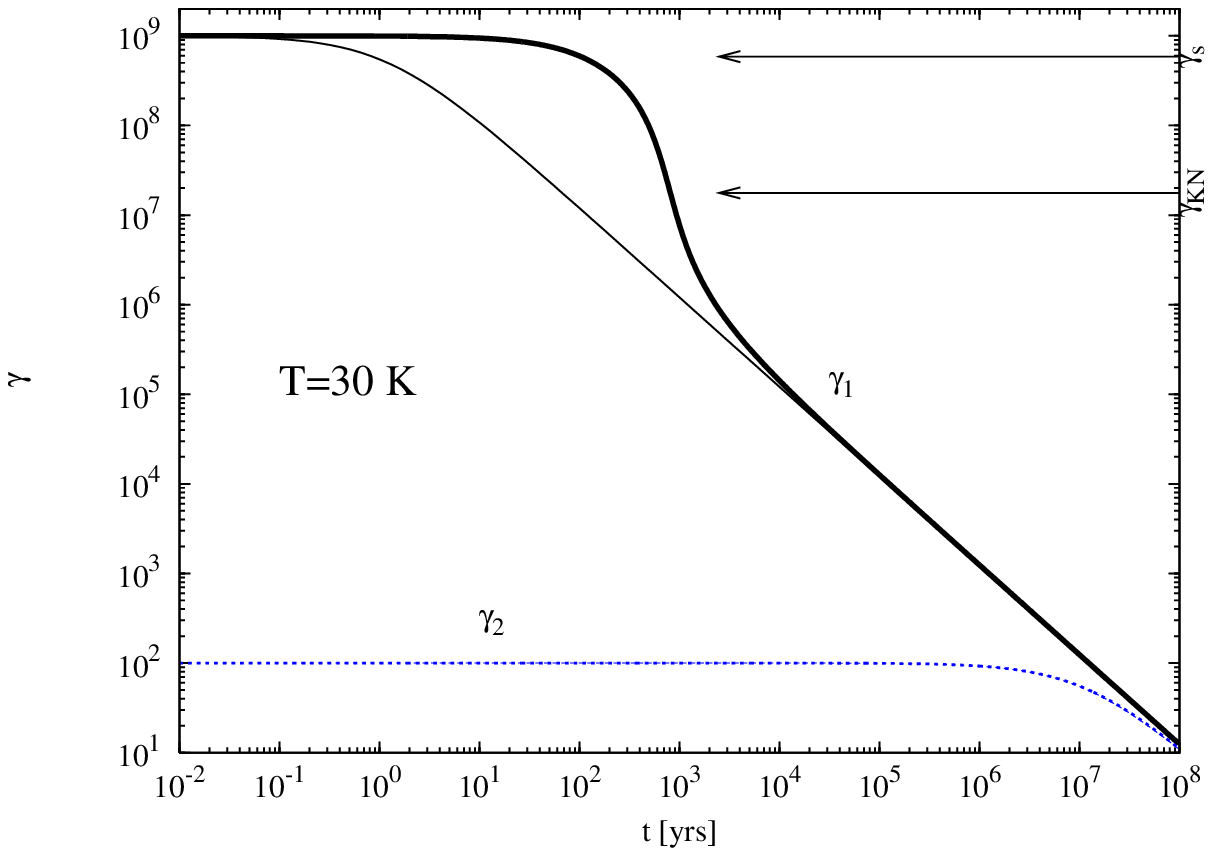}
\caption{\label{fig:g0g1} The quantities 
$\gamma_1$
and $\gamma_2$ defined in Eqs.~(\ref{gamma1}) and (\ref{gamma2})
as functions of time for magnetic fields and 
target photon fields of the same 
energy density but for two different photon temperatures: 
$T=30\,000\,$K (left panel)
and $T=30\,$K (right panel). In both panels, $\gamma_{\rm s}$ 
(transition from synchrotron to KN cooling) 
and $\gamma_{\rm KN}$ (transition from KN to Thomson) are indicated by
horizontal arrows.
For comparison, the thin lines indicate the evolution of $\gamma_1$ 
and $\gamma_2$ in the Thomson limit
(for $\gamma_2$, the cooling proceeds at all times in the Thomson regime, since
$\gmin$ is chosen to be 100; therefore, the two curves are 
superposed).}
\end{figure*}

The cooling timescale for electrons 
$\tau_{\rm cool}(\gamma)=\gamma/\dot \gamma$ for the
various energy loss mechanisms, assuming an environment where inverse
Compton cooling in the Klein-Nishina limit is of importance, is shown
in Fig.~\ref{fig1}. The parameters used are
$n_{\rm e}=1$~cm$^{-3}$, $u_0=500$~eV~cm$^{-3}$, $T=30\,000$~K, and
$B=10~\mu$G, as expected in an
environment dominated by hot and young stars which provide a 
relatively high
temperature target photon field.  The cooling time for Coulomb
scattering and Bremsstrahlung losses is calculated assuming a
fully ionised gas. This is a self-consistent assumption taking into
account the high temperature and energy density
of the ambient photon field, which is sufficient
to ionize neutral hydrogen.  The cooling time is dominated by different 
mechanisms in different energy ranges. At the lowest energies,
Coulomb scattering (\lq\lq ionisation loss\rq\rq) 
is the fastest energy loss mechanism, whereas above
approximately $100\,$MeV, inverse Compton scattering in the Thomson
regime takes over. For electron energies above a few GeV, the inverse
Compton scattering becomes less efficient, because of the drop
in the cross section. At energies larger than $1\,$TeV, inverse Compton
scattering is less efficient than synchrotron radiation, which remains 
the dominant mechanism at higher
energies. Although the transition energies
described depend on the choice of the parameters 
($u_0$, $n_{\rm e}$, $T$, and $B$), the qualitative picture remains the same.
It is interesting to compare the approximation for $F_{\rm KN}(\gm)$ with
the \lq\lq exact\rq\rq\ calculation shown in Fig.~\ref{fig1}
(using the expression for $f_{\rm KN}(\gamma,\epso)$ given by
\citet{2005MNRAS.364.1488M}). The
approximation slightly overpredicts $\tau_{\rm cool}$, but is fairly
accurate up to very high energies ($\sim10\%$ at $E\simeq 10^{14}\,$eV, 
corresponding to $\gm\epsilon_{\rm eff}\simeq 10^4$). 
However, as one can readily deduce from
Fig.~\ref{fig1}, the approximation for $F_{\rm KN}(\gm)$ becomes inaccurate
only in the regime where synchrotron losses are likely to
dominate. Even in binary systems, where the radiation field
density can be substantially higher than the value assumed in
Fig.~\ref{fig1}, the ratio of $u_0/u_B$ is unlikely to be larger than
$O(1000)$ and, therefore, electrons with $\gm\epsilon_{\rm eff}\ga 10^4$ cool 
predominantly through synchrotron radiation.


\begin{table}
\begin{center}
{\small 
\begin{tabular}{l|cc}
Photon distribution & $\bt$ & $\epsilon_{\rm eff}$  \\
\hline 
&&\\
Planck, temperature $T$ & $-3/2$ & $2.8kT/\melec c^2$   \\
&&\\
power-law, $\al_0<-0.5$ & $-3/2$ & $\epsmax$   \\
&&\\
power-law, $-0.5<\al_0<1$ & $\al_0-1$& $ \epsmax$   \\
&&\\
power-law, $\al_0>1$ & $-3/2$ & $ \epsmin$  
\end{tabular}
}
\end{center}
\caption{The values of the parameters $\bt$ and $\epsilon_{\rm eff}$ 
(see Eq.~\ref{FKNapprox}) for
two different photon energy distributions: a (possibly diluted) Planckian
of temperature $T$, or, equivalently a mono-energetic distribution 
with energy 
$\epso = 2.8kT/\melec c^2$, and a power-law photon field with spectral index 
$\al_0 \equiv -\diff \ln u_{\epso}/\diff \ln \epso$. For the case where 
$-0.5<\al_0<1$, it is further
assumed that $\epsmin/\epsmax \ll 1$. All approximations are valid for 
$4\gm\epsilon_{\rm eff} < 10^4$.}
\label{table1}
\end{table}
\subsection{Time dependent solutions for the particle distribution}
For astrophysical environments with dilute plasma and fields of high
energy density, 
the most relevant energy loss mechanisms for high energy electrons are
synchrotron and inverse Compton radiation.  Therefore, one may safely
discard in this case the last two terms in Eq.~(\ref{gammadottotal})
which then becomes: 
\beq 
\dot{\gm}=b_s\gm^2\left[1+\bis F_{\rm KN}(\gm)\right],
\label{gammadot}
\eeq
\noindent
where $\bis \equiv b_{\rm iC}/b_s=u_0/u_{\rm B}$ denotes 
the ratio of the photon field total energy density to the magnetic
field energy density.
Using Eqs.~(\ref{FKNapprox}) and (\ref{gammadot}) we may write 
this in the general form:
\beq
\dot{\gm}=b_s\gm^2\left[1+\bis 
\left(1+4\gm\epsilon_{\rm eff}\right)^\bt\right],
\label{gammadotgen}
\eeq
\noindent
where $\bt$ and $\epsilon_{\rm eff}$ are chosen from 
Table~\ref{table1} for the appropriate target photon field. 
Taking into account Eqs.~(\ref{escrate}) and (\ref{gammadotgen}),
the functions describing cooling time and escape probability
given in Eq.~(\ref{f1f2})
can be expressed as
\bea 
\tau(\gm',\gm)&=&
\int_\gm^{\gm'}\!\!\frac{\diff \gm''}{b_s {\gm''}^2 
\left[1+\bis\left(1+4\gm''\epsilon_{\rm eff}\right)^{\bt}\right]}
\label{taubohm}
\\
\lm(\gm',\gm)&=& \int_\gm^{\gm'}\!\!\frac{b_{\rm es} \diff \gm''}
{\gm''\left[1+\bis\left(1+4\gm''\epsilon_{\rm eff}\right)^{\bt}\right]} 
\label{lmbohm}
\eea where the notation $b_{\rm es} \equiv b_{\rm esc}/b_s $ has been
used. Note that for $\gm'\rightarrow\infty$, 
$\tau(\gm',\gm)$ remains finite,
whereas $\lm(\gm',\gm)$ diverges. Physically, this is because
the synchrotron cooling rate, which goes as $\gm^2$ dominates at 
high Lorentz factor, so that the time
taken to cool to $\gm$ from an arbitrarily large $\gm'$ is finite.
However, if, as in the case of Bohm diffusion, 
the mean free path of the electrons increases in proportion to 
$\gm$, the probability of escape during this finite cooling time 
tends to unity as $\gm'\rightarrow\infty$.

In Appendix~\ref{appendixa}, we give the analytical form of the
functions $\tau(\gm',\gm)$ and $\lm(\gm',\gm)$ 
assuming that $\nu_{\rm esc}(\gm)
\propto \gm$ (see Eq~\ref{escrate}). For a power-law type photon
energy spectrum with spectral index $-0.5<\al_0<1$, these functions
are given in closed form only for the special 
cases $\al_0=0$ and $\al_0=1/2$, although 
a numerical evaluation is straightforward
using Eqs~(\ref{definitionT}) and (\ref{definitionL}).
The analytical expressions have the advantage that the solution of 
Eq.~(\ref{parN}) is reduced to the single quadrature given in 
Eqs~(\ref{sol2N}) and (\ref{sol2N2})
or (\ref{sol3N}).

In the Thomson limit for Compton scattering $F_{KN} \simeq 1$.
Therefore, assuming $\nu_{\rm esc}=\besc\gamma$, one has 
$\beta=0$ in Eqs.~(\ref{taubohm}) and (\ref{lmbohm}), 
and the functions $\tau(\gm',\gm)$ and $\lambda(\gm',\gm)$
can be written down explicitly: 
\bea
\dot{\gm}&=&(b_s+b_{\rm iC})\gm^2 \equiv b \gm^2
\\
\tau(\gm',\gm)&=&\frac{1}{b}\left(\frac{1}{\gm}-\frac{1}{\gm'}\right)
\label{tauthomsonlimit}
\\
\noalign{\hbox{and}}
\lm(\gm',\gm)&=&\frac{\besc}{b} \ln\left(\frac{\gm'}{\gm}\right)
\eea
The solution of Eq.~(\ref{parN}) follows from 
Eqs.~(\ref{sol2N}) and (\ref{sol2N2}).
For $\gm<\gamma_1(t)=1/(bt)$: 
\bea 
N(\gm, t) &=& \frac{1}{b
  \gm^{(2b-\besc)/b}}
\Bigl[
\int\limits_\gm^{\gamma_0}\!\!\diff \gm'{\gm'}^{-\besc/b}
\nonumber\\
\lefteqn{%
Q \left(\gm',t-\frac{1}{b\gm}+\frac{1}{b\gm'}\right)
H \left(t-\frac{1}{b\gm}+\frac{1}{b\gm'}\right)\Bigr]}&&
\nn \\
&+& \frac{1}{b\gm^2}\left(\frac{\gm}{\gm_0}\right)^{\besc/b}N(\gm_0,0)
\enspace,
\label{sol1Ncompton}
\eea
where $\gamma_0=\gamma/(1-bt\gamma)$
and, for $\gm>1/(bt)$:
\bea 
N(\gm, t) &=& \frac{1}{b
  \gm^{(2b-\besc)/b}}
\int\limits_\gm^{\infty}\!\!\diff \gm'{\gm'}^{-\besc/b}
Q \left(\gm',t-\frac{1}{b\gm}+\frac{1}{b\gm'}\right)
\enspace.
\label{sol1Ncomptonb}
\eea
The solution for a continuous power-law injection
with finite energy spectrum is 
given in Appendix~\ref{appendixb}.

In the case where the escape rate does not depend on the particle
energy, $\lm(\gm',\gm)=\tau(\gm',\gm) /t_{\rm esc}$, where $t_{\rm esc} =
1/\nu_{\rm esc}=\textrm{const}.$ 
Then, in the Thomson limit of Compton scattering $\tau(\gm'\gm)$ is 
given, as before, by Eq.~(\ref{tauthomsonlimit}), and from 
Eq.~(\ref{sol2N}) one finds: 
\bea 
N(\gm, t) &=&
\frac{1}{b \gm^2}  \textrm{e}^{-t_{\rm cool}/t_{\rm esc}} 
\Bigl\lbrace
\int\limits_\gm^{\gm_0}\!\!\diff \gm'\textrm{e}^{1/b\gm't_{\rm esc}}
\nn \\
\lefteqn{%
Q \bigl[\gm',t-(b\gm)^{-1}+(b\gm')^{-1}\bigr]
H \bigl[t-(b\gm)^{-1}+(b\gm')^{-1}\bigr]\Bigr\rbrace}&&
\nn \\
&+& \frac{1}{b \gm^2}  \textrm{e}^{-t_{\rm cool}/t_{\rm esc}}
N(\gm_0,0)\textrm{e}^{1/bt_{\rm esc}\gm_0}
\enspace,
\label{sol2Ncompton}
\eea
for $\gm>1/(bt)$, and for $\gm<1/(bt)$:
\bea
N(\gm, t) &=&
\frac{1}{b \gm^2}  \textrm{e}^{-t_{\rm cool}/t_{\rm esc}}\nn \\
&\times& 
\int\limits_\gm^{\infty}\!\!\diff \gm'\textrm{e}^{1/b\gm't_{\rm esc}}
Q \bigl[\gm',t-(b\gm)^{-1}+(b\gm')^{-1}\bigr]
\enspace,
\eea
where the cooling time $t_{\rm cool}(\gm):=
\tau(\infty,\gm)=1/b \gm$ is
the time taken for an electron to cool from infinite 
Lorentz factor to Lorentz factor $\gm$.  
For $N(\gm,0)=0$, with 
a constant mono-energetic injection spectrum, $Q(\gm,t)
=Q_0 H(t)\dl(\gm-\gm_i)$, 
and for $\gm_i>\gm>\gm_i/\left(1+bt\gm_i\right)$, 
Eq.~(\ref{sol2Ncompton}) yields a steady-state solution:
\beq 
N(\gm,t)=\frac{Q_0}{b \gm^2} \textrm{e}^{-t_{\rm
    cool}/t_{\rm esc}+1/\gm_i t_{\rm esc}}.
\label{solkardashev}
\eeq
\noindent
This solution displays a cut-off towards lower Lorentz factors
at $t_{\rm cool}=t_{\rm esc}$ 
as has been pointed out by \citet{1962AZh....39..393K}.

\section{Numerical results}
\label{sec:numerical}
\subsection{Comparison with the steady state solution}
\label{subsec:ss}
Ignoring escape losses, Eq.~\ref{parN} 
in the steady state limit, $\partial N/\partial t \rarrow 0$, 
becomes: 
\beq 
{\diff \over \diff \gm} \{\dot{\gm}
N(\gm)\} +Q(\gm) = 0 \enspace .  
\eeq 
which is easily integrated to give: 
\beq 
N(\gm)=-\dot{\gm}^{-1}
\int\limits_{\gamma}^{\infty} Q(\gm')\diff\gm'.
\label{ss-lim}
\eeq In this case it is straightforward to incorporate the general
term for $\dot\gm$ given by Eq.~(\ref{gammadottotal}).  In
Fig.~\ref{fig2}, we compare this solution with the time dependent
solution obtained by using Eqs.~(\ref{sol2N}) and (\ref{sol2N2}) with
the approximate cooling term given in Eq.~(\ref{gammadotgen}).  We
take $t=10^8$~years, which is sufficiently large to allow a steady
state to be achieved, and assume no particles are present initially,
$N(\gamma,0)=0$.  In this figure, escape is neglected, $\bes=0$, and
the ratio of the energy density of the photon field to that of the
magnetic field is $\bis=200$. The remaining parameters are the same as
in Fig.~\ref{fig1} Note that here, and in the following, we restrict
our calculations to the case of a thermal distribution of target
photons. The qualitative picture is similar to the case of a photon
field following a power-law.  

For $\gamma\la 100$, Coulomb losses
harden the spectrum to $p-1=1$. Since the time-dependent solution is
based upon a \lq\lq simplified\rq\rq\ energy-loss term given in
Eq.~(\ref{gammadot}) neglecting the Coulomb energy losses, the two
solutions deviate in this range.
Between $\gamma\simeq 300$ and $\gamma\simeq \gamma_{\rm
  KN}=(4~\epsilon_0)^{-1}\approx 18\,000 (T/30\,000~{\rm K})^{-1}$,
electrons cool predominantly via inverse Compton scattering in the
Thomson limit. For values of $\gamma>\gamma_{\rm KN}$, the drop in the
inverse Compton scattering cross section in the Klein-Nishina regime
leads to a spectral hardening until synchrotron losses become dominant
for $\gamma\ga \gamma_{\rm s}=(\bis^{2/3}-1)/(4\epsilon_0)\approx
6\times10^5 (\bis/200)^{2/3} (T/30\,000~{\rm K})^{-1}$.  The positions
of $\gamma_{\rm KN}$ and $\gamma_{\rm s}$ are indicated in Fig.~\ref{fig2}.
The slight deviation observed for $\gamma_{\rm KN} \la \gamma \la
\gamma_{\rm s}$ is the result of the approximation used for $F_{\rm KN}$
given in Eqn.~(\ref{fKNapprox}). From Fig.~\ref{fig2}, it is obvious
that (a) the approximation used to take the energy dependence of the
inverse Compton scattering cross section into account follows quite
well the exact calculation and (b) neglecting Coulomb losses mainly
affects the low energy part of the spectrum.
\subsection{Temporal evolution}

The temporal evolution of $N(\gamma,t)$ for an arbitrary injection
function $Q(\gamma,t)$ can now be calculated.  In the following we
investigate the impact of the
relevant parameters ($b_{s},\bis,\bes,T$). 
In order to demonstrate the effect of KN cooling on the temporal
evolution, we plot in Fig.~\ref{fig:g0g1} the evolution of $\gamma_1$
and $\gamma_2$ as a function of time for two different temperatures
($T=30\,000$~K and $T=30$~K) which are representative for an
environment dominated by hot stars or by re-processed emission from
warm dust. For comparison, the evolution of $\gamma_1$ and $\gamma_2$
in the Thomson limit is depicted by thin lines. In both cases, the
parameters have been chosen similar to the ones used above
($B=10~\mu$G, $u_0=500~$eV~cm$^{-3}$ corresponding to $\bis=200$,
$\bes=0$) with $\gmin=100$ and $\gmax=10^9$.  The evolution of
$\gamma_2$ is dominated by slow cooling in the Thomson limit and
therefore, the two lines are identical. For the high temperature
environment (Fig.~\ref{fig:g0g1}a), $\gamma_1$ cools via synchrotron
radiation until roughly 1~Myr, when the 
energy
losses in the KN limit take over. Within a few hundred thousand years,
the transition to Thomson-limit cooling takes place when $\gamma_1
< \gamma_{\rm KN}$. Qualitatively, the evolution of $\gamma_1$ for a
cooler photon field proceeds in a similar way (Fig.~\ref{fig:g0g1}b).
It is, however, important to note that the respective transitions occur
much earlier. Whereas for $T=30\,000$~K, synchrotron cooling prevails
for about $1\,$Myr, for $T=30\,$K, Compton 
cooling in the KN regime is reached 
within a few hundred years. 

\begin{figure}
\includegraphics[width=\linewidth]{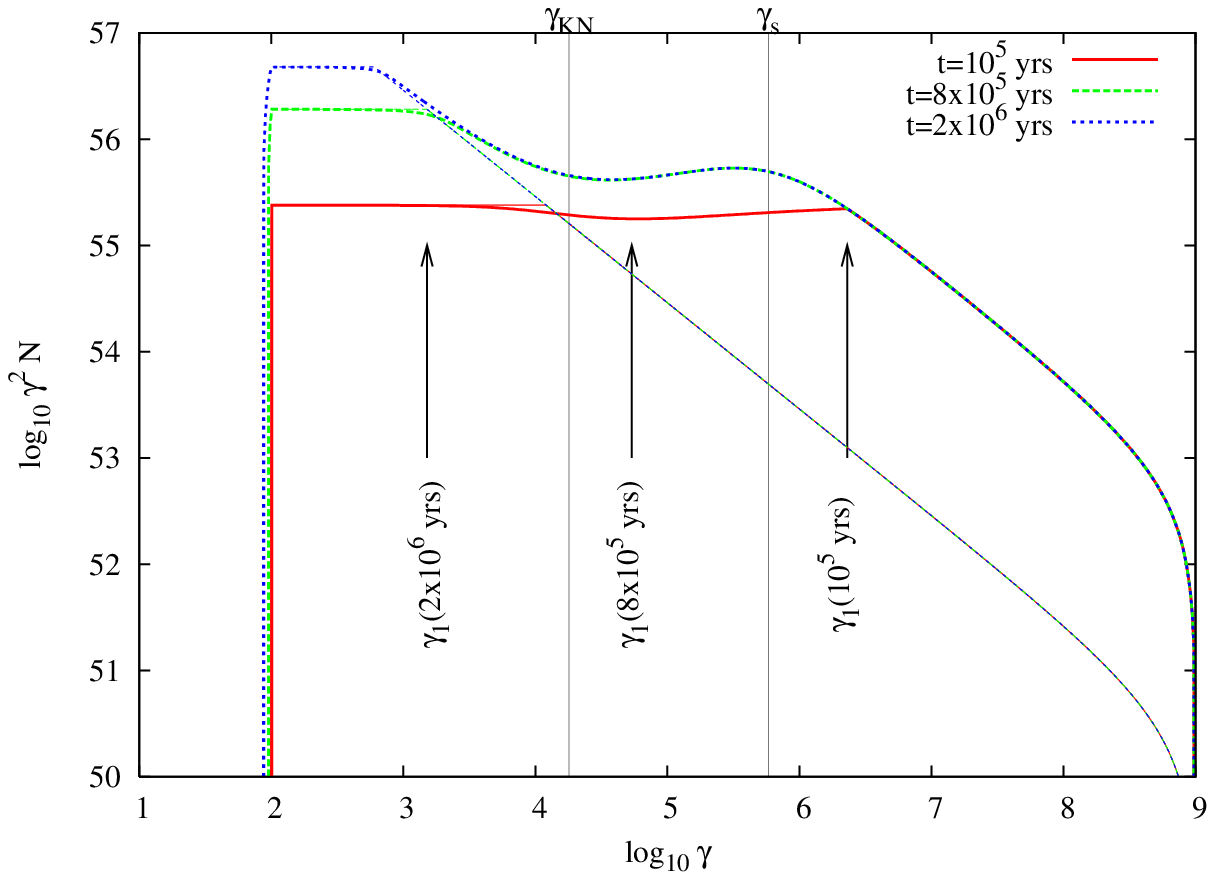}
\caption{\label{fig:demo} The solution $N(\gamma,t)$ obtained
  including the effects of KN cooling (thick lines) and in the Thomson
  limit (see Appendix C) for three different times chosen such that
  $\gamma_1>\gamma_{\rm s}$ ($t=10^5$~yrs), $\gamma_{\rm KN} < \gamma_1<
  \gamma_{\rm s}$ ($t=8\times 10^5$~yrs), and $\gamma_1 < \gamma_{\rm KN}$
  ($t=2\times 10^6$~yrs). The parameters used are identical to the
  ones used e.g. in Fig.~\ref{fig:g0g1}a. The positions of $\gamma_1$
  at different times as well as of $\gamma_{\rm KN}$ and $\gamma_{\rm s}$
  are indicated by arrows and vertical lines respectively.}
\end{figure}

The effect on the temporal evolution of a particle distribution which
is partially cooling via inverse Compton radiation in the KN limit,
assuming continuous power-law
electron injection with $p=2$ between $\gmin=10^2$ and $\gmax=10^9$,
neglecting escape losses ($\bes=0$) and with 
the environmental parameters described above
is
shown in Fig.~\ref{fig:demo}.
The solutions $N(\gamma,t)$ for
$t=10^5$, $8\times 10^5$, and $2\times 10^6$ years, 
weighted with $\gamma^2$ so 
that the injected spectrum appears constant, are compared with
solutions found 
using expressions for the cooling valid in 
the Thomson limit (see Appendix~\ref{appendixb}), which are 
depicted as thin lines. 

The solutions for $t=10^5$~yrs (see also Fig.~\ref{fig:g0g1}a)
correspond to the case $\gamma_1>\gamma_{s}$ (the position of
$\gamma_1$ is marked by an arrow in Fig.~\ref{fig:demo}).  The
solution taking KN effects into account shows a break such that
$N\propto \gamma^{-(p+1)}$ for $\gamma>\gamma_1$ while for smaller
$\gamma_{\rm KN}<\gamma<\gamma_{\rm s}$, a slight deviation from the $p=2$
injection power-law can be observed which marks the onset of rapid
energy loss in the KN regime. In the solution obtained in the Thomson
limit, the break occurs at a smaller value of $\gamma_1$ (see also
Fig.~\ref{fig:g0g1}a) which marks the transition from the uncooled
part of the spectrum to the part which has already suffered from
cooling. At the solution at a later time ($t=8\times 10^5$~yrs), when
$\gamma_{\rm KN}< \gamma_1<\gamma_{\rm s}$, the flattened spectral shape
presented earlier in the steady state solution (see Fig.~\ref{fig2}),
has already fully developed as a consequence of the large $\dot
\gamma$ in the KN regime. The solution obtained in the Thomson limit
and that obtained taking KN effects into account, agree with each
other for $\gamma<\gamma_{\rm KN}$, which marks the transition to
cooling in the Thomson limit. Finally, for $t=2\times 10^6$~yrs,
$\gamma_1<\gamma_{\rm KN}$, and the further temporal evolution is well
described in the Thomson limit.

\begin{figure}
\includegraphics[width=\linewidth]{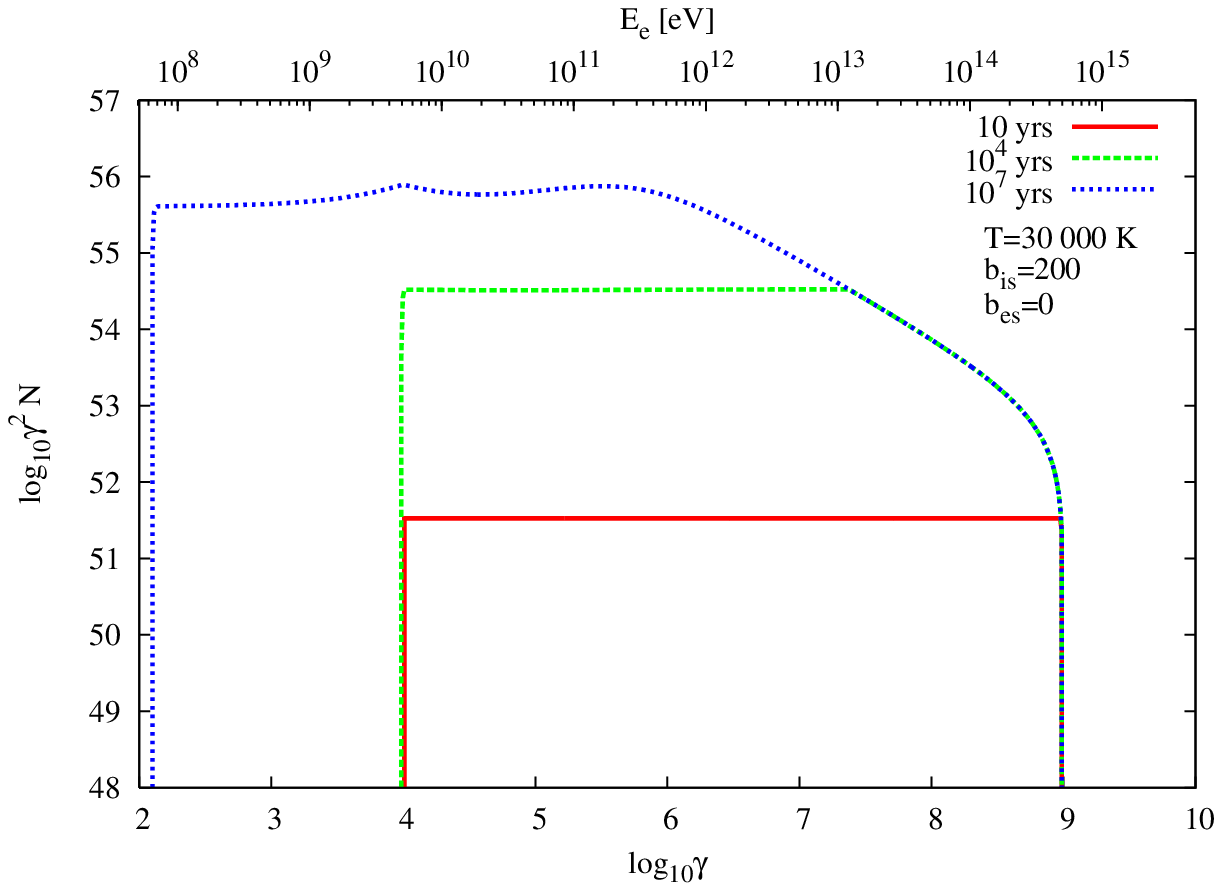}
\caption{\label{figevolve} Temporal evolution of an electron
  distribution injected with $p=2$ between $\gamma_{\rm min}=10^4$ and
  $\gamma_{\rm max}=10^9$ in a hot ($T=30\,000$~K) photon field.  The
  evolution is shown for three distinct times:
  $t=10$, $10^4$, and $10^7\,$years. The energy density of the 
photon field is
  200 times that of the
  magnetic field ($\bis=200$) and escape losses are neglected
  ($\bes=0$).\label{fig:evolve}}
\end{figure}

In the next example, we consider a higher value of $\gmin=10^4$
while keeping the same values for all other parameters as in the
examples above.  Fig.~\ref{fig:evolve}, shows
$N(\gamma,t)$ for three different times: $t=10$, $10^4$, and $10^7$~years.  
The shape of the spectrum evolves, as expected, mainly
due to synchrotron cooling for $t\la 10^6$~years.  
At later times (e.g., $t=10^7$~yrs), 
the spectrum shows a smooth transition between $p=2$
and $p+1=3$ at $\gamma\approx \gamma_{\rm s}$. For smaller values of
$\gamma$, the spectrum remains almost flat, showing a slight peak at
the position of $\gmin$. The overall shape of this evolved
spectrum could, in principle, mimic the case of an uncooled spectrum up
to TeV energies (depending on the value of $\bis$) with a softening at
$\gamma\approx \gamma_{\rm s}$ resembling a much younger source.
Remarkably, for $p=2$, 
the effect of cooling in KN is almost hidden when $\gmin\approx
\gamma_{\rm KN}$.  For higher values of $\bis$, a hard particle
spectrum can be retained up to even higher values of $\gamma$ than those
shown here (see also next example).

\begin{figure}
\includegraphics[width=\linewidth]{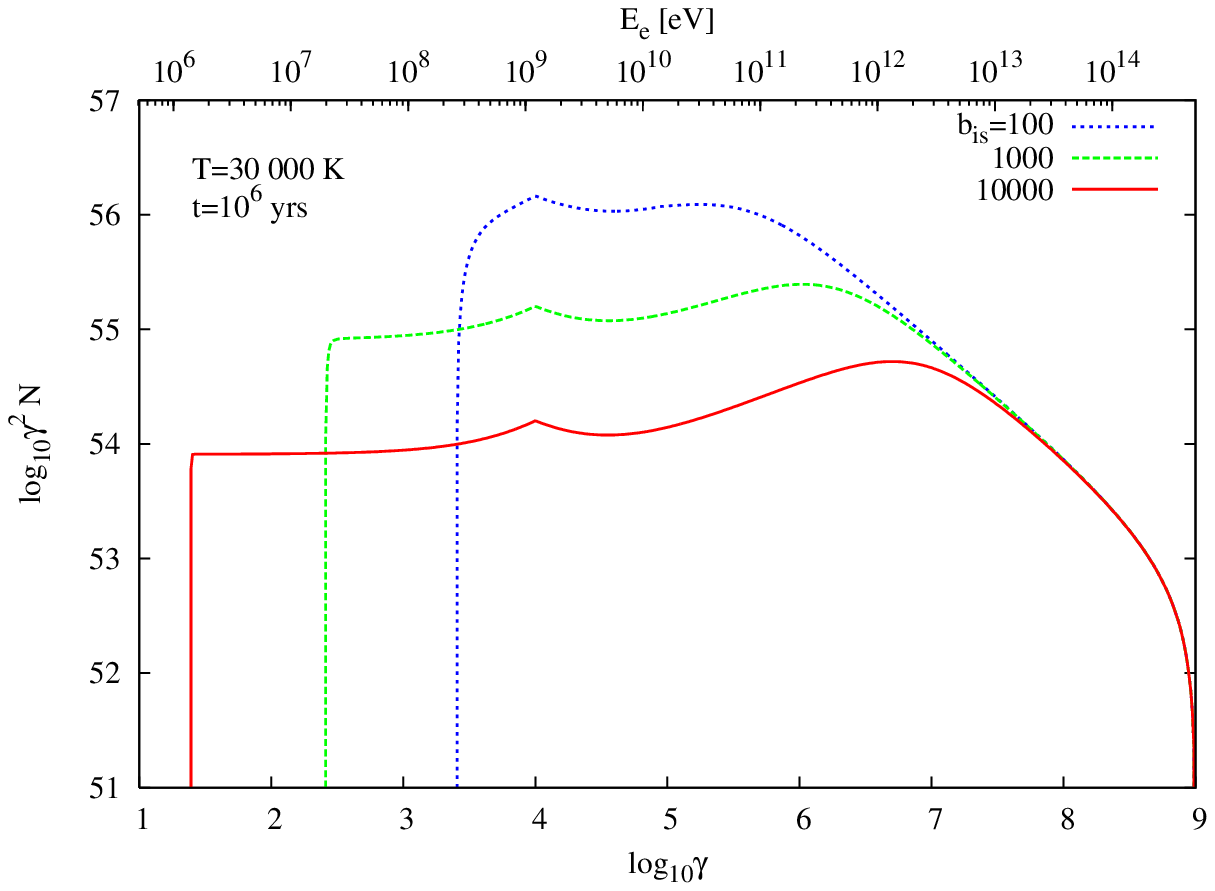}
\caption{\label{figbis} For a similar set of parameters as described
  in Fig.~\ref{fig:evolve} and a fixed time ($t=10^6$~yrs) the effect
  of varying $\bis=100,1000,10^4$ is demonstrated. }
\end{figure}

Changing the ratio of the photon and magnetic field energy densities 
($\bis$) at fixed photon temperature influences
the location of the 
transition between inverse Compton and synchrotron cooling. In
Fig.~\ref{figbis}, the effect of varying $\bis$ from values of 100 up
to 10$^4$ is demonstrated. Increasing $\bis$ leads to more rapid
cooling and a more pronounced spectral hardening as a result of
inverse Compton cooling in the KN limit.

\begin{figure}
\includegraphics[width=\linewidth]{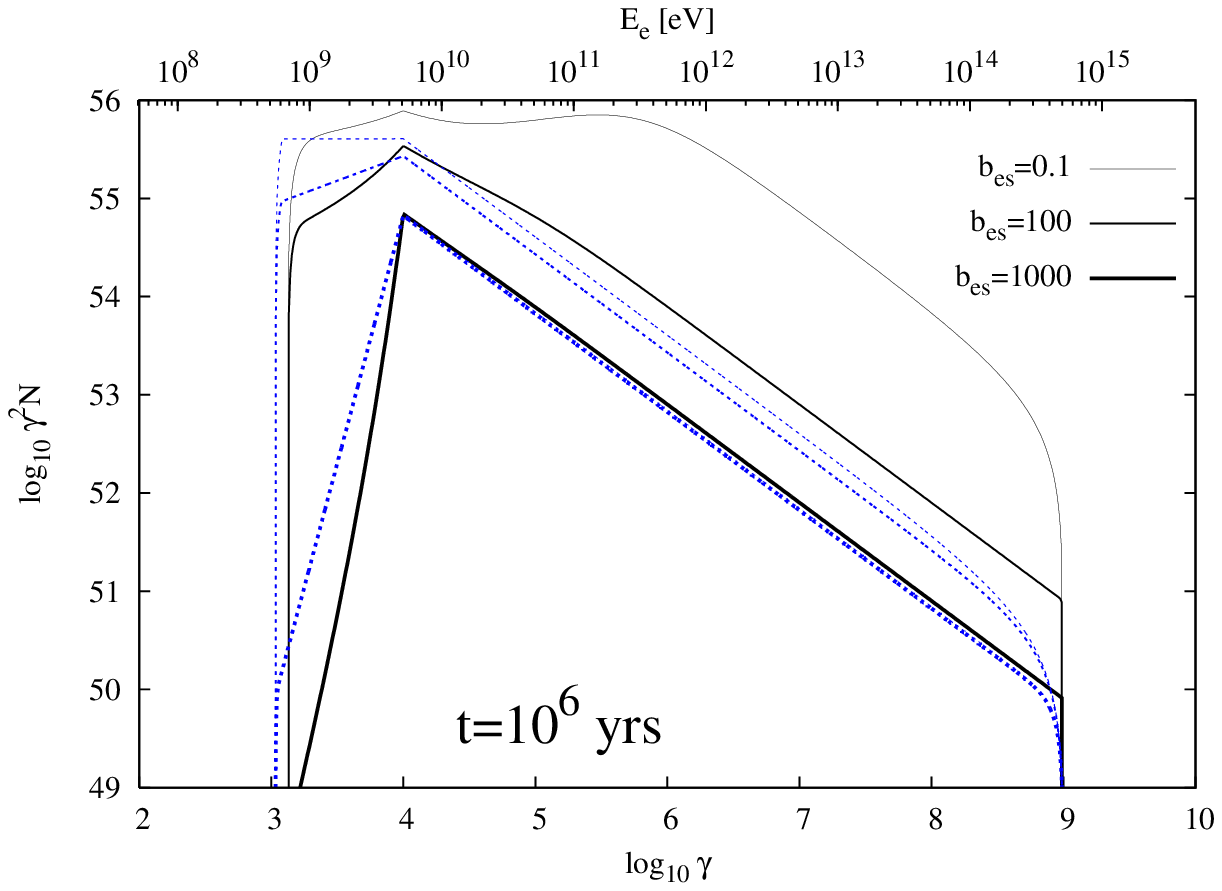}
\caption{\label{figbes} The effect of escape losses is demonstrated
  for three different values of $\bes=0.1,100,1000$ (thin, medium, and
  heavy lines).  The time has been fixed to $1$~Myr. The injection
  rate as well as the other parameters are identical to the cases
  discussed in the previous figures. The solid lines are obtained
  taking KN effects into account while for comparison the analytical
  solution for the Thomson limit is shown with dashed lines.  }
\end{figure}

Finally, we consider the effect of particle escape via diffusive
particle transport leaving a sphere with radius $R$. In
Fig.~\ref{figbes}, the solution $N(\gamma,t)$ is shown for values of
$\bes=0.1$, $100$, and $1000$ at $t=10^6$ yrs (solid lines). 
For comparison, the solution of Appendix~\ref{appendixb} 
in the Thomson limit is also shown (dashed lines). 
All other parameters are the same
as in the previous example. 
As expected, the effect of escape losses leads to
a reduction of total particle number. For
$\bes=0.1$ escape losses are negligible, whereas for $\bes=100$ and
$\bes=1000$, escape losses modify the solution considerably in
comparison to the case of small $\bes$. The solution taking KN effects
into account shows a stronger relative suppression of particle numbers
for $\gamma>\gamma_{\rm S}$ than in the Thomson limit. This is a
consequence of the fact that, in the KN case, radiative cooling is
dominated by synchrotron losses for $\gamma>\gamma_{\rm S}$ which
implies that escape losses are relatively more important 
($\lambda(\gm',\gm)\sim\besc/b_{s}$) 
than in the Thomson limit, where $\lambda(\gm',\gm)\sim\besc/b$ with
$b=b_s+b_{\rm iC}$. In general, when escape losses are important, 
the spectral features characteristic for
cooling in the KN limit are suppressed, and the solution 
resembles that found in 
the Thomson limit. Asymptotically, for very large
$\bes\gg b$, radiative cooling is negligible and the two solutions
converge.

\section{Application to a stellar association: Westerlund 2}
\label{sec:applic}
 \bfig
  \includegraphics[width=\linewidth]{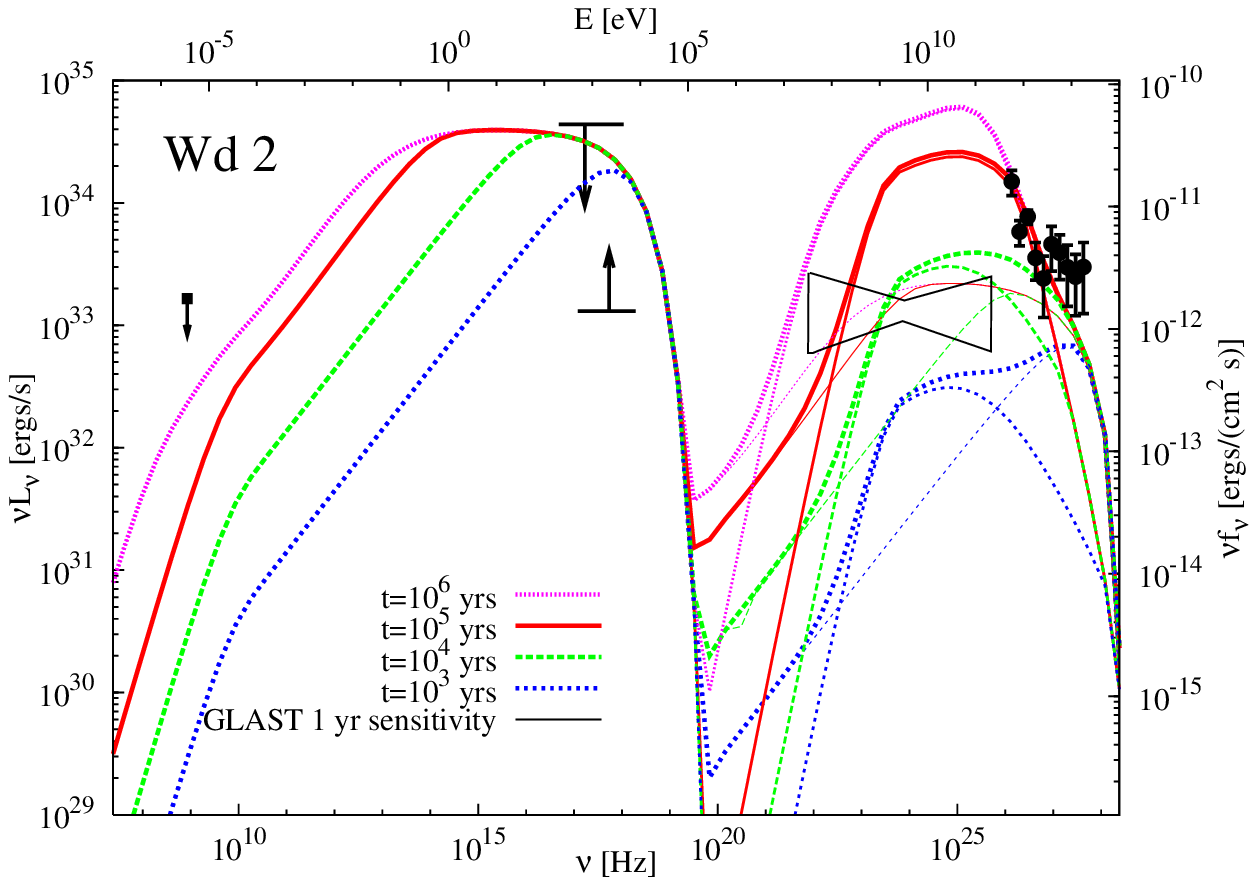}
  \caption{\label{fig:wd2}Broadband spectral energy distribution of
    Westerlund 2 (Wd 2): the radio point represents the integrated 843
    MHz emission of the core given by \citet{1997A&A...317..563W} and
    is considered to be an upper limit of a possible non-thermal
    component. The X-ray lower limit is based upon the flux observed
    in the spectral hard tail detected with Chandra
    \citep{2005xrrc.procE3.04T}.  We consider this detection a lower
    limit because of the limited field of view of the Chandra
    telescope. The upper limit in the X-ray band is based upon
    Einstein IPC observations possibly overestimating the total flux
    because of limited spatial resolution
    \citep{1987ApJ...322..349G}. The VHE data is taken from
    \citet{2007A&A...467.1075A} and the GLAST sensitivity is indicated
    for a one year observation. The model curves are derived for four
    different times assuming a continuous injection of a power-law
    with $p=2$ between $\gmin=10^4$ and $\gmax=3\times 10^{8}$ in a
    spherical region with 8.8~pc diameter. The total power injected is
    $7\times 10^{35}$ ergs~s$^{-1}$ assuming a distance of
    $D=2.8$~kpc. The escape loss, synchrotron, and inverse Compton
    cooling are taken into account assuming $B=10~\mu$G,
    $u_0=500~$eV~cm$^{-3}$, and $T=30\,000$~K. For the inverse Compton
    emission, the heavy line indicates the sum of the emissivity from
    the stellar light (medium curve) allowing for KN effects as well
    as from the cosmic microwave background (thin curve).}
 \efig
 The discovery of VHE gamma-ray emission from young stellar
 associations like Westerlund 2 (Wd-2), Cyg OB2, and Berkeley 87 is
 providing first clear evidence that particle acceleration can occur
 in these systems. While the origin of the emission and the nature of
 the accelerated particles is not clear, it has been suggested that
 gamma-rays from stellar associations could be the decay product of
 neutral mesons in inelastic scattering of nuclei with the ambient
 medium \citep{2004ApJ...601L..75T,2007arXiv0704.3517B} or
 due to the excitation of giant resonances of nuclei by 
Doppler boosted UV photons \citep{2007PhRvL..98l1101A}. 
 Here, we calculate the broad band spectral energy distribution of
 Wd-2 at different stages of its evolution.  The distance to this
 system is not very well constrained \citep{2004ApJS..154..322C}. 
 A recent estimate places
 Wd-2 at $2.8\,$kpc, with an age of $2.0\pm 0.3$ Myrs, and
 a total mass of $10^4\,M_\odot$ \citep{2007A&A...466..137A}. 
 While
 this distance estimate is based mainly {on near infra-red 
 photometry and colors}, the spectroscopic investigation of the cluster and the
 light-curve of the binary system WR~20a has provided a larger
 distance estimate of $8.0\pm1.4$~kpc \citep{2007A&A...463..981R}.
 {Another recent distance estimate is based upon the possible association of
 RCW~49 with a giant molecular cloud with a mass
 of $7.5\times 10^5~M_\odot$ at a kinematic distance of $6.0\pm1.0$~kpc
 \citep{2007arXiv0707.1360D}}
 For
 our calculations we adopt a distance of $D=2.8$~kpc and note that the
 larger distance would affect the total energetics, requiring a higher injection rate.

 The angular extension, $0.18\pm0.02^\circ$
 of the VHE gamma-ray emission region
 \citep{2007A&A...467.1075A} 
implies a radial extension of $8.8~{\rm pc}\cdot (D/2.8~{\rm
   kpc})$. The total luminosity of the member stars of O-type has been
 estimated to be $L_{\rm bol}\approx 3\times 10^{40}$~ergs~s$^{-1}$
 \citep{2007A&A...463..981R}.  However, it is well known that the
 number of UV photons emitted by the early cluster member stars is not
 sufficient to power the HII region RCW 49 in which Wd-2 is embedded.
 Based upon the estimates of \citet{2007A&A...463..981R}, only
 $\approx 20$~\% of the UV photons required to power RCW~49 can be
 attributed to known O stars. Therefore, the assumed luminosity of the
 early type stars is certainly a lower limit and the total luminosity
 could in fact reach values beyond $10^{41}$~ergs~s$^{-1}$.  It is
 noteworthy, that the population of later type stars shows a rather
 flat projected spatial distribution even beyond a distance of 
$4\,$pc
 \citep{2007A&A...466..137A} which would lead to a higher average
 energy density of the photon field.  In addition to the stellar
 photons, re-processed emission from the dust present in Wd-2 and its
 surrounding will provide a contribution to the seed photon field for
 inverse Compton scattering which is however not included here and
 left for a more
 detailed modelling.

 Taking into account just the total power emitted by the early type
 stars in the cluster, the average energy density of the hot photon
 field at a distance $R$ corresponding to the extension of the VHE
 gamma-ray source can be estimated as:
\begin{eqnarray}
  u_0 &=& 13\,500~\frac{\rm eV}{\rm cm^3} 
\left(\frac{L_{\rm bol}}{3\times 10^{40}~{\rm ergs~s}^{-1}}\right)
\left(\frac{R}{\rm pc}\right)^{-2}.
\end{eqnarray}
Based upon the extension of the VHE source of $8.8$~pc,
$u_0=174$~eV~cm$^{-3}$. However, given the presence of additional
(diffuse) UV photons implied by the ionisation state of RCW~49, we
choose $u_0=500$~eV~cm$^{-3}$ as the average energy density and a
temperature of $T=30\,000$~K to characterise the 
target photon field. 
The magnetic field is set to
$10~\mu$G, placing it in approximate equipartition 
with the thermal energy density of the gas, 
assuming a source
radius of $8.8$~pc. 
The injection rate $Q(\gamma)$ is assumed to be
constant in time with a total power of $7\times 10^{35}~$ergs~s$^{-1}$
following a power-law with $p=2$ between $\gmin=10^4$ and
$\gmax=3\times 10^8$.  
The required injection power can be compared, for example, with
the total power injected by the stellar winds driven 
by the early type stars (mainly the Wolf-Rayet stars 
WR~20a\&b) present in Wd-2 which has
been estimated to be $5\times 10^{37}$~ergs~s$^{-1}$ 
\citep{2007A&A...463..981R}.
If the emitting electrons are accelerated by these stellar winds, 
they must take
roughly $1\,$\% of the kinetic power of the bulk flow.

In hadronic scenarios, on the other hand, the required efficiency 
is generally much larger, 
because of the relatively long cooling times involved. For example, 
inelastic proton-proton scattering
producing $\pi^0$, 
operates on a timescale of 
$t(pp\rightarrow \pi^0+\ldots)\approx 1.5\times 10^8 n_1^{-1}$~yrs 
($n=n_1~\mathrm{cm}^{-3}$ denotes the ambient medium density), which
is larger than the age of the accelerator, assuming this
is limited to the maximum
time that massive stars 
drive strong and fast stellar winds in their Wolf-Rayet phase 
(this lasts approximately, 
$t_{WR}\approx 5\times 10^5$~yrs, see also below). 
This implies that under most conditions, the efficiency required in
hadronic models is close or larger than unity. 
A similar conclusion can be drawn 
for the case of gamma-ray production 
through photo-excitation of heavy nuclei, where
the efficiency of acceleration of exclusively iron nuclei has to 
reach values in excess of $8~\%$ 
\citep{2007arXiv0706.0517A}.

The electron distribution $N(\gamma,t)$ is computed for four different
times $t=10^3$, $10^4$, $10^5$, 
and $10^6$ years. The synchrotron and inverse Compton
emissivities are calculated following the approach described in Appendix~\ref{appendixc}
and the resulting broad band spectral
energy distribution (SED) is presented in Fig.~\ref{fig:wd2}. In
addition to the UV photons from the early type stars, CMB photons are,
of course, also present with an energy density $u_{\rm CMB}=0.26~{\rm
  eV~cm}^{-3}$. Although the energy loss 
due to inverse Compton scattering on
these photons is negligible, they are nevertheless quite
important for the total hard gamma-ray emission. This is 
because the
inverse Compton scattering off UV photons is strongly suppressed, 
leading for sufficiently large $\gamma$ to emissivities which are
smaller than the contribution from scattering with CMB photons. We
therefore include in the inverse Compton emissivity calculation the
CMB photon field and show in Fig.~\ref{fig:wd2}, the total (heavy line
width) inverse Compton emission as well as the contribution from UV
photons (medium line width) and from CMB photons (thin line).  It is
interesting to note how the inverse Compton emission from the CMB
photons dominates for young sources. With increasing age,  
the UV-related component 
at about $1\,$TeV 
increases steadily until, after $\approx
10^6\,$years --- roughly the age of this stellar cluster --- 
it exceeds the CMB-related contribution.
This effect is due to the accumulation of  
electrons of Lorentz factor ($\gamma\approx 10^6$) within the source
(see Fig.~\ref{figevolve}). 

The model SED is compared in Fig.\ref{fig:wd2} 
with VHE measurements 
\citep{2007A&A...467.1075A} and with X-ray measurements 
\citet{2005xrrc.procE3.04T} taken with Chandra (indicated as a lower
limit) and earlier measurements taken with the Einstein X-ray
telescope \citep{1987ApJ...322..349G} (indicated as an upper
limit). Whereas the limited field of view of the Chandra observations
may underestimate the total extended X-ray emission, the Einstein
observations may overestimate the diffuse flux, because of the
limited spatial resolution which does not allow the 
subtraction of the point
sources present. In this figure we also show 
the total integrated radio flux of $210\,$Jy at
843~MHz. This we consider as an upper limit which should not be exceeded
by the non-thermal radio flux produced by the energetic electrons 
\citep{1997A&A...317..563W}.
Overall, the model SED is in good agreement with the observations. The
VHE energy spectrum is well reproduced for $t>10^5\,$years, and 
the constraints from
X-ray and radio measurements are not violated. 
It is interesting to note 
that for stars with $M>25~M_\odot$, the Wolf-Rayet phase is expected to 
last up to roughly $t_{WR}\approx 5\times 10^5$ 
years \citep{1987A&A...182..243M}, which is consistent with the injection 
time required to match the data in this model.
In contrast, an age of $t>10^5$ yrs would be too long for 
acceleration in a supernova remnant, a hypothesis which
also has to contend with the fact that,  
so far, no indication for the presence of a supernova remnant
associated with Wd-2 has been reported.  

The predicted X-ray flux
is quite close to the upper 
bound imposed by the Einstein
measurements. However, it should be noted that 
this prediction is sensitive to the energy density in the target photon 
field --- 
an increase by a factor of 2 in the
energy density in the seed photon field (either from the late type
stars and/or emission from dust) reduces the required injection
power and, consequently, the predicted X-ray flux is also reduced
by a factor of 2. 
We also indicate in Fig.~\ref{fig:wd2} the one year
flux sensitivity of the upcoming GLAST gamma-ray
mission\footnote{taken from 
{\tt www-glast.slac.stan.edu},
official GLAST homepage}. 
Our calculations predict that GLAST
should easily detect the source.  The combined energy spectrum of
GLAST and H.E.S.S. would provide a broad band energy spectrum which in
turn might allow one to infer, for example, the age of the source in the model
scenario suggested here.
 
 \section{Conclusions}
 \label{sec:conclusions}
We have presented a method for computing the time-dependent electron 
distribution in the continuous energy loss limit  
for an arbitrary source
function, under the assumptions of 
\begin{itemize}
\item
an isotropic electron distribution
\item
negligible ionization and
 Bremsstrahlung losses, 
\item
time-independent energy losses by 
synchrotron radiation in a homogeneous
magnetic field and 
\item
by inverse Compton scattering off 
a spatially 
homogeneous mono-energetic or power-law energy
distribution of target photons, 
\item
an escape probability proportional to the particle energy, 
as expected for diffusion in the Bohm limit.
\end{itemize}
The key innovation of the model is that it utilises 
a simplified expression introduced by \citet{2005MNRAS.363..954M}
for the inverse Compton scattering rate (including KN-effects) to
construct explicit analytic expressions (see Appendix~\ref{appendixa})
for the integrated cooling time
($\tau(\gm',\gm)$, Eq.~\ref{taudefinition})) 
and the integrated escape function
($\lambda(\gm',\gm)$, Eq.~(\ref{f1f2})). This enables 
the electron distribution to be computed in a single quadrature. 

The extension of this approach to include different energy dependencies of 
the escape probability would be relatively straightforward. However, 
the treatment of multiple and/or time-dependent target photon fields 
would probably require numerical evaluation of the functions
$\tau$ and $\lambda$. 

We have used this method to present an in-depth discussion of the
properties of cooling electron distributions, and have identified 
features characteristic of cooling by inverse Compton scattering:
\begin{enumerate}
\item The effect of \lq\lq spectral ageing\rq\rq\ observed in the Thomson limit
  is strongly modified. While in the Thomson limit, a well-defined
  spectral break $\gamma_1$ changes position with time (cooling
  break), this spectral break is not evident anymore for
  $\gamma_1<\gamma_{\rm s}$, when KN effects are taken into account
  (see Fig.~\ref{fig:demo}).
\item For $\gamma_{\rm KN}<\gamma<\gamma_{s}$, a spectral hardening is
  observed as a result of cooling in the KN regime (see
  Figs.~\ref{fig2},\ref{fig:demo}).
\item When choosing $\gmin\approx \gamma_{\rm KN}$, the evolved
  spectrum shows a smooth transition at $\gamma_{s}$ resembling a
  cooling break. However, this transition is stationary in time and
  may be mistakenly interpreted as a cooling break (see
  e.g. Figs.~\ref{figevolve}-\ref{figbis}).
\item The effect of particle escape modifies the observed energy
  spectra stronger for $\gamma>\gamma_{s}$ than for smaller values of
  $\gamma$ which can lead to energy spectra resembling more the ones
  obtained in the Thomson limit (see Fig.~\ref{figbes}).
\end{enumerate}

We have illustrated our technique by applying it to high energy
observations of the open stellar cluster Westerlund~2, and calculating
the broad band emission spectrum.  Comparing our results with the
available observations, we have identified a consistent parameters
set: $\bis=200$, $B=10~\mu$G, $T=30\,000$~K, total injected power in
electrons $7\times 10^{35}$~ergs~s$^{-1}$, distance=2.8~kpc,
$R=8.8$~pc, $\gmin=10^4$, $\gmax=3\times 10^8$, $p=2$.  Although the
uncertainty on the distance affects the values of some parameters
(eg. $R, \bis$, and total power), it is interesting to note that the
suggested model of an electron accelerator embedded in the intense
stellar radiation field which has been injecting a power-law type
spectrum for the past few $10^5$~yrs, can account naturally for the
observed gamma-ray spectrum, without violating the constraints derived
from X-ray and radio observations. 
Furthermore, the rather long injection time
favours acceleration in stellar winds of, for example, Wolf-Rayet stars 
over acceleration in supernova remnants. A leptonic scenario of gamma-ray
production requires only a moderate fraction ($\sim 1$~\%) 
of the kinetic
power of the stellar winds to be channelled into the 
acceleration of a non-thermal particle
population. Hadronic models, on the other hand, need a rather high 
ambient gas density
in order to keep the required efficiency below unity.
The model presented here can be improved
further by including a more realistic mixture of different temperature
photon fields with different spatial extensions as observed in the
vicinity of Wd-2 (including a dust component and later-type stars
present in the system). However, it already makes testable order of
magnitude predictions of the flux in the range accessible to the GLAST
experiment.

 
\begin{acknowledgements}
DH and KM acknowledge the support of the Eberhard Karls Universit\"at
T\"ubingen.
This research has made use of NASA's Astrophysics Data System. 
\end{acknowledgements}

\newpage
\onecolumn
\appendix
\section{Characteristic cooling and escape functions}
\label{appendixa}
The integrals in Eqs~(\ref{taubohm}) and (\ref{lmbohm}) 
can be rewritten as
\bea
\tau(\gm',\gm)&=&
\frac{8\epsilon_{\rm eff}}{b_s}
\int_{\sqrt{1+4\epsilon_{\rm eff}\gm}}^{\sqrt{1+4\epsilon_{\rm eff}\gm'}}
\frac{\om^{1-2\beta}\,\diff\om}
{\left(\om^2-1\right)^2\left(\om^{-2\beta}+\bis\right)}
\label{definitionT}
\\
&=&\frac{8\epsilon_{\rm eff}}{b_s}
\left[ T_{\beta}\left(\sqrt{1+4\epsilon_{\rm eff}\gm'}\right)
-T_{\beta}\left(\sqrt{1+4\epsilon_{\rm eff}\gm}\right)\right]
\nonumber
\eea
and
\bea
\lm(\gm',\gm)&=&
2b_{\rm es}
\int_{\sqrt{1+4\epsilon_{\rm eff}\gm}}^{\sqrt{1+4\epsilon_{\rm eff}\gm'}}
\frac{\om^{1-2\beta}\,\diff\om}
{\left(\om^2-1\right)\left(\om^{-2\beta}+\bis\right)}
\label{definitionL}
\\
&=& 2b_{\rm es}
\left[ \Lambda_{\beta}\left(\sqrt{1+4\epsilon_{\rm eff}\gm'}\right)
-\Lambda_{\beta}\left(\sqrt{1+4\epsilon_{\rm eff}\gm}\right)\right]
\nonumber
\eea
where the indefinite integrals $T_{\beta}(\om)$ and $\Lambda_{\beta}(\om)$ 
can be expressed in terms of elementary functions
for (negative) integer and half-integer values of $\beta$.

In the case of black-body target photons, as well as 
for power-law photons with index $\alpha_0<-0.5$, and 
$\alpha_0>1$, one has $\beta=-3/2$, and finds,
for $\bis\ne1$:
\bea
T_{-3/2}(\om) &=& a^{(1)}_{-3/2}\ln(\om_1^2-\bis^{1/3}\om_1+\bis^{2/3})
+a^{(2)}_{-3/2}\arctan\left(\frac{2\om_1-\bis^{1/3}}{\sqrt{3} \bis^{1/3}}\right)
+a^{(3)}_{-3/2}\ln(\om_1+\bis^{1/3})+a^{(4)}_{-3/2}\ln(\om_1+1) \nn \\
&+& a^{(5)}_{-3/2}\ln(\om_1-1) +a^{(6)}_{-3/2}(\om_1-1)^{-1}+a^{(7)}_{-3/2}(\om_1+1)^{-1},
\label{F1bispl}
\eea
\noindent
where\\
\begin{tabular}{lll}
& & \\
$ a^{(1)}_{-3/2} (\bis)=-\bis^{2/3}(2\bis^2+\bis^{8/3}-6\bis^{4/3}+2\bis^{2/3}+1)/6(\bis^2-1)^2$, &   &
$ a^{(5)}_{-3/2} (\bis)=3\bis/4(\bis+1)^2$, \\
$ a^{(2)}_{-3/2} (\bis)=-\bis^{2/3}(2\bis^2-\bis^{8/3}-2\bis^{2/3}+1)/\sqrt{3}(\bis^2-1)^2$, &  &
$ a^{(6)}_{-3/2} (\bis)=-1/4(\bis+1)$, \\
$ a^{(3)}_{-3/2} (\bis)=\bis^{2/3}(2\bis^2+\bis^{8/3}+3\bis^{4/3}+2\bis^{2/3}+1)/3(\bis^2-1)^2$, &  &
$ a^{(7)}_{-3/2} (\bis)=-1/4(\bis-1)$ \\
$ a^{(4)}_{-3/2} (\bis)=-3\bis/4(\bis-1)^2$, &          & \\
&  &  \\
\end{tabular}\\
and, for $\bis=1$:
\beq
T_{-3/2}(\om)=-\frac{1}{9}\ln(\om^2-\om+1)+\frac{5}{144}\ln(\om+1)
+\frac{3}{16}\ln(\om-1)+
\frac{1}{24}\frac{\om^2-7\om-6}{(\om+1)^2(\om-1)}
\enspace.
\label{F11pl}
\eeq
For the indefinite integral $\Lambda_{-3/2}(\om)$ one finds, for $\bis\ne1$: 
\bea
\Lambda_{-3/2}(\om) &=& c^{(1)}_{-3/2}\ln(\om^2-\bis^{1/3}\om+\bis^{2/3})
+c^{(2)}_{-3/2}\arctan\left(\frac{2\om-\bis^{1/3}}{\sqrt{3} \bis^{1/3}}\right)
+c^{(3)}_{-3/2}\ln(\om+\bis^{1/3})+c^{(4)}_{-3/2}\ln(\om+1) \nn \\
&+& c^{(5)}_{-3/2}\ln(\om-1),
\label{F2bispl}
\eea
where\\
\begin{tabular}{lll}
& & \\
$ c^{(1)}_{-3/2} (\bis)=[(2\bis^{2/3}-1)\bis^{4/3}-\bis^{2/3})]/6(\bis^2-1)$, &   &
$ c^{(4)}_{-3/2} (\bis)=-1/2(\bis-1)$, \\
$ c^{(2)}_{-3/2} (\bis)=(\bis^{4/3}-\bis^{2/3})/\sqrt{3}(\bis^2-1)$, &  &
$ c^{(5)}_{-3/2} (\bis)=1/2(\bis+1)$ \\
$ c^{(3)}_{-3/2} (\bis)= [(\bis^{2/3}+1)\bis^{4/3}+\bis^{2/3})]/3(\bis^2-1)$,&  & \\
&  &  \\
\end{tabular}\\
and, for $\bis=1$:
\beq
\Lambda_{-3/2}(\om)=\frac{1}{6}\ln(\om^2-\om+1)
+\frac{1}{3\sqrt{3}}\arctan\left(\frac{2\om-1}{\sqrt{3}}\right)
+\frac{5}{12}\ln(\om+1)+\frac{1}{4}\ln(\om-1)
+\frac{1}{6}(\om+1)^{-1}
\enspace.
\label{F21pl}
\eeq

In the case of a power-law distribution of target photons with
$-0.5<\alpha_0<1$, closed form expressions of the integrals can be
found for the special cases $\alpha_0=0$ and $\alpha_0=1/2$,
corresponding to $\beta=-1$ and $\beta=-1/2$, respectively. For $\beta=-1$:
\beq
T_{-1}(\om)= -\frac{\bis}{2(1+\bis)^2}\ln(\om^2+\bis)+
\frac{\bis}{2(1+\bis)^2}\ln(\om^2-1)
 -\frac{1}{2(\bis+1)} (\om^2-1)^{-1},
\label{F10}
\eeq
\beq
\Lambda_{-1}(\om)=\frac{\bis}{2(1+\bis)}\ln(\om^2+\bis)+
\frac{1}{2(1+\bis)}\ln(\om^2-1) 
\enspace.
\label{F20}
\eeq
whereas for $\beta=-1/2$ one has for $\bis\ne1$:
\beq
T_{-1/2}(\om)=
a^{(1)}_{-1/2}\ln(\om+\bis)+
a^{(2)}_{-1/2}\ln(\om+1)+
a^{(3)}_{-1/2}\ln(\om-1) + 
a^{(4)}_{-1/2}(\om+1)^{-1}+
a^{(5)}_{-1/2}(\om-1)^{-1}
\enspace,
\label{F1bis05}
\eeq
\noindent
where\\
\begin{tabular}{lll}
& & \\
$ a^{(1)}_{-1/2} (\bis)=\bis^2/(\bis^2-1)^2$, &   &
$ a^{(4)}_{-1/2} (\bis)=-1/4(\bis-1)$, \\
$ a^{(2)}_{-1/2} (\bis)=-\bis/4(\bis-1)^2$, &  &
$ a^{(5)}_{-1/2} (\bis)=-1/4(\bis+1)$ \\
$ a^{(3)}_{-1/2} (\bis)= \bis/4(\bis+1)^2$,&  & \\
&  &  \\
\end{tabular}\\
and, for $\bis=1$:
\beq
T_{-1/2}(\om)=-\frac{1}{16}\ln(\om+1)+\frac{1}{16}\ln(\om-1)+
\frac{1}{8}\frac{\om^2-3\om-2}
{(\om+1)^2(\om-1)}
\enspace.
\label{F1105}
\eeq
\noindent
Similarly, for the function $\Lambda_{-1/2}$ we find, for 
$\bis\ne1$:
\beq
\Lambda_{-1/2}(\om)=
c^{(1)}_{-1/2}\ln(\om+\bis)+
c^{(2)}_{-1/2}\ln(\om+1)+
c^{(3)}_{-1/2}\ln(\om-1) 
\enspace,
\label{F2bis05}
\eeq
where\\
\begin{tabular}{lll}
&&\\
$c^{(1)}_{-1/2} (\bis)=\bis^2/(\bis^2-1)$,&&
$c^{(3)}_{-1/2} (\bis)=1/2(\bis+1)$,\\
$c^{(2)}_{-1/2} (\bis)=-1/2(\bis-1)$&&\\
&&\\
\end{tabular}\\ 
and, finally, for $\bis=1$:
\beq
\Lambda_{-1/2}(\om)=\frac{3}{4}\ln(\om+1)+\frac{1}{4}\ln(\om-1)+
\frac{1}{2}(\om+1)^{-1}
\enspace.
\label{F2105}
\eeq

\section{}
\label{appendixb}
For the case where accelerated particles start to inject from the central source at time $t_0$ with a power-law type finite energy spectrum:
\beq
Q=
\left\lbrace \begin{array}{ll}
Q_0 \gm^{-p}& \textrm{for $\gmin \le \gamma \le \gmax$ } \\
        &       \\
0 & \textrm{otherwise}
\end{array}
\right. 
\enspace,
\label{inj2}
\eeq
\noindent
undergoing synchrotron and inverse Compton losses in the Thomson regime: $\dot{\gm}=(b_s+b_{\rm iC})\gm^2 \equiv b \gm^2$, and escaping from the system (with radius $R$) at a rate $\nu_{\rm esc} = b_{\rm esc} \gm$ (see \S~\ref{subrates}), the solution of Eq.~(\ref{parN}) splits into two branches,
depending on whether the maximum injected energy, $\gmax$ has had enough time to ``cool down'' to a 
value $\gm_1$ being smaller or larger than the minimum injected energy, $\gmin$. \\
\\
If $\gmin < \gm_1$, then the solution reads:

\begin{equation}
 N(\gamma,t)= \left\{ \begin{array}{ll}
0,& \mbox{$\gamma< \gm_2$}\\
        &       \\
K{\gamma}^{-(p+1)}
\left\lbrace \left({\gamma \over \gmin}\right)^{q}-
\left[1-b \gamma(t-t_0)\right]^{q}\right\rbrace, &
\mbox{$\gm_2 \le \gamma \le \gmin$}\\
        &       \\
K{\gamma}^{-(p+1)}
\{1-[1-b \gamma(t-t_0)]^{q} \}, &
\mbox{$\gmin< \gamma < \gm_1 $}\\
        &       \\
K{\gamma}^{-(p+1)}\left[1-({\gamma \over
\gmax})^{q}\right], &
\mbox{$\gm_1 \le \gamma \le \gmax$}\\
        &       \\
0, & \mbox{$ \gamma > \gmax$}
\end{array}
\right.
\label{csol1}
\end{equation}
\noindent
If $\gmin > \gm_1$, one gets:

\begin{equation}
 N(\gamma,t)= \left\{ \begin{array}{ll}
0,& \mbox{$\gamma< \gm_2$}\\
        &       \\
K{\gamma}^{-s}
\left\lbrace \gmin^{-q}-
\left[\gm^{-1}-b (t-t_0)\right]^{q}\right\rbrace, &
\mbox{$\gm_2 \le \gamma \le \gm_1$}\\
        &       \\
K{\gamma}^{-s}
(\gmin^{-q} -\gmax^{-q}), &
\mbox{$\gm_1< \gamma < \gmin $}\\
        &       \\
K{\gamma}^{-(p+1)}\left[1-({\gamma \over
\gmax})^{q}\right], &
\mbox{$\gmin \le \gamma \le \gmax$}\\
        &       \\
0, & \mbox{$ \gamma > \gmax$}
\end{array}
\right.
\label{csol2}
\end{equation}
\noindent
where $q=p-1+\besc/b$, $K={Q_0/ bq}$, $s=2-\besc/b$ and the values of $\gm_1$ and $\gm_2$ are given by Eqs~(\ref{gamma1}) and (\ref{gamma2}), respectively.

\section{Inverse Compton and synchrotron emissivities}
\label{appendixc}
The spectrum of photons scattered by an energetic electron  from an isotropic photon 
gas which follows a differential photon number density 
$\diff n=n(\epsilon)\diff \epsilon$ has been 
derived e.g. by \citet{1970RvMP...42..237B} in the head-on collision approximation and is given by their 
Eqn.2.48:
\begin{eqnarray}
\frac{\diff N_{\gamma,\epsilon}}{\diff t \diff \epsilon_1}&=& 
\frac{3\sigma_T m c^3}{\gamma} \frac{n(\epsilon) \diff \epsilon}{\epsilon}\cdot f(q)
\end{eqnarray}
with 
\begin{eqnarray}
 f(q) &=& 2q\ln q + (1+2q)(1-q)+\frac{1}{2}\frac{(4\epsilon \gamma/mc^2)^2}{1+4q\epsilon\gamma/mc^2 }(1-q)
\end{eqnarray}
and 
\begin{eqnarray}
 q&=&  \frac{\epsilon_1}{4\epsilon \gamma/mc^2 (1-\epsilon_1)}.
\end{eqnarray}
The total inverse Compton spectrum from a distribution of electrons $dN=N(\gamma)d\gamma$ follows from the integration
over $\epsilon$ and $\gamma$:
\begin{eqnarray}
 \frac{\diff N_{IC}}{\diff \epsilon_1 \diff t} &=& \int \diff \epsilon \int \diff \gamma N(\gamma) \frac{\diff N_{\gamma,\epsilon}}{\diff t \diff \epsilon_1}.
\end{eqnarray}

 The emitted synchrotron power per unit frequency interval emitted by a single electron 
 with pitch angle $\alpha$ and magnetic field $B$ is given e.g. by \citet{1964ocr..book.....G}:
 \begin{eqnarray}
  j(\nu) &=& \frac{\sqrt{3} e^3 B \sin\alpha}{m c^2} F\left( \frac{\nu}{\nu_c} \right)
 \end{eqnarray}
 with the critical frequency:
 \begin{eqnarray}
  \nu_c &=& \gamma^2 \frac{3 eB \sin\alpha}{4\pi m c}
 \end{eqnarray}
 and
 \begin{eqnarray}
 F(x) &=& x \int\limits_x^\infty \diff t K_{5/3}(t).
 \end{eqnarray}
A simple approximation of $F(x)$ has been used for the calculations presented in this paper (see e.g. 
\citet{1980gbs..bookR....M}):
\begin{eqnarray}
 F(x) &=& 1.85 x^{1/3} \exp(-x).
\end{eqnarray}
This approximation provides a relative accuracy better than 1 per cent in the region
 of the maximum at $x\sim 0.29$, and still reasonable accuracy in the broad range of $0.1\le x\le 10$.
The luminosity per unit frequency interval from an electron distribution 
$\diff N = N(\gamma)\diff \gamma$ is
calculated by integration over $\gamma$:
\begin{eqnarray}
 J(\nu) &=& \int \diff \gamma N(\gamma) j(\nu).
\end{eqnarray}
We assume an isotropic magnetic field such that $\sin \alpha=\sqrt{2/3}$.


%
%
%

\twocolumn
\bibliography{jk,libros}
\bibliographystyle{aa}
\end{document}